\documentclass[onecolumn,a4paper,accepted=2022-09-12]{quantumarticle}
\usepackage{amsmath}
\usepackage{amssymb}
\usepackage{amsthm}
\usepackage{amsfonts}
\usepackage[caption=false]{subfig}
\usepackage[colorlinks]{hyperref}
\usepackage[all]{hypcap}
\usepackage{tikz}
\usepackage{relsize}
\usepackage{color,soul}
\usepackage[utf8]{inputenc}
\usepackage{capt-of}
\usepackage{mathtools}
\usepackage[numbers,sort&compress]{natbib}
\usepackage{float}
\usepackage[section]{placeins}
\usepackage{listings}
\usepackage[T1]{fontenc}  
\usetikzlibrary{decorations.pathreplacing}
\usepackage{braket}

\DeclareFixedFont{\ttb}{T1}{txtt}{bx}{n}{4}
\DeclareFixedFont{\ttm}{T1}{txtt}{m}{n}{4}
\definecolor{deepblue}{rgb}{0,0,0.5}
\definecolor{deepred}{rgb}{0.6,0,0}
\definecolor{deepgreen}{rgb}{0,0.5,0}
\newcommand\cppstyle{\lstset{
language=C++,
basicstyle=\ttm,
otherkeywords={uint8_t, __m256i, size_t, ASSERT_TRUE, EXPECT_TRUE, TEST, BENCHMARK},
keywordstyle=\ttb\color{deepblue},
emphstyle=\ttb\color{deepblue},
stringstyle=\color{deepgreen},
commentstyle=\fontfamily{txtt}\selectfont\color{gray},
showstringspaces=false,
literate={*}{{\char42}}1
         {-}{{\char45}}1
}}
\lstnewenvironment{cpp}[1][]
{\cppstyle\lstset{#1}}{}

\newcommand\pythonstyle{\lstset{
language=python,
basicstyle=\ttm,
morekeywords={assert,as,echo},
keywordstyle=\ttb\color{deepblue},
emphstyle=\ttb\color{deepblue},
stringstyle=\color{deepgreen},
commentstyle=\fontfamily{txtt}\selectfont\color{gray},
showstringspaces=false,
literate={*}{{\char42}}1
         {-}{{\char45}}1
}}
\lstnewenvironment{python}[1][]
{\pythonstyle\lstset{#1}}{}

\lstdefinestyle{stimcircuit}{
    language=python,
    basicstyle=\fontsize{4}{4}\selectfont\ttfamily,
    upquote=true,
    stepnumber=1,
    numbersep=8pt,
    showstringspaces=false,
    breaklines=true,
    frame=single,
    aboveskip=1.5em,
    belowskip=1.5em,
    commentstyle=\color{gray},
    classoffset=1,
    morekeywords={DETECTOR,OBSERVABLE_INCLUDE,rec},
    keywordstyle=\color{deepgreen},
    classoffset=2,
    morekeywords={H,R,MPP,M},
    keywordstyle=\ttb\color{deepblue},
    classoffset=3,
    morekeywords={X_ERROR,DEPOLARIZE2,DEPOLARIZE1},
    keywordstyle=\color{red},
    classoffset=4,
    morekeywords={TICK,SHIFT_COORDS,QUBIT_COORDS},
    keywordstyle=\color{gray}
}

\renewcommand\thesection{\arabic{section}}

\theoremstyle{definition}

\theoremstyle{definition}

\theoremstyle{definition}

\newcommand{\eq}[1]{\hyperref[eq:#1]{Equation~\ref*{eq:#1}}}
\renewcommand{\sec}[1]{\hyperref[sec:#1]{Section~\ref*{sec:#1}}}
\DeclareRobustCommand{\app}[1]{\hyperref[app:#1]{Appendix~\ref*{app:#1}}}
\newcommand{\fig}[1]{\hyperref[fig:#1]{Figure~\ref*{fig:#1}}}
\newcommand{\tbl}[1]{\hyperref[tbl:#1]{Table~\ref*{tbl:#1}}}
\newcommand{\theoremref}[1]{\hyperref[theorem:#1]{Theorem~\ref*{theorem:#1}}}
\newcommand{\definitionref}[1]{\hyperref[definition:#1]{Definition~\ref*{definition:#1}}}

\begin{document}
\title{Benchmarking the Planar Honeycomb Code}

\date{\today}
\author{Craig Gidney}
\email{craig.gidney@gmail.com}
\affiliation{Google Quantum AI, Santa Barbara, California 93117, USA}

\author{Michael Newman}
\email{mgnewman@google.com}
\affiliation{Google Quantum AI, Santa Barbara, California 93117, USA}

\author{Matt McEwen}
\email{mmcewen@google.com}
\affiliation{Google Quantum AI, Santa Barbara, California 93117, USA}
\affiliation{University of California, Santa Barbara, 93106, USA}

\begin{abstract}
We improve the planar honeycomb code by describing boundaries that need no additional physical connectivity, and by optimizing the shape of the qubit patch.
We then benchmark the code using Monte Carlo sampling to estimate logical error rates and derive metrics including thresholds, lambdas, and teraquop footprints.
We determine that the planar honeycomb code can create a logical qubit with one-in-a-trillion logical error rates using 7000 physical qubits at a 0.1\% gate-level error rate (or 900 physical qubits given native two-qubit parity measurements).
Our results cement the honeycomb code as a promising candidate for two-dimensional qubit architectures with sparse connectivity.
\end{abstract}

\emph{The source code that was written, the exact noisy circuits that were sampled, and the statistics that were collected as part of this paper are available at \href{https://doi.org/10.5281/zenodo.7072889}{doi.org/10.5281/zenodo.7072889}~\cite{gidneyhoneycombdata2022}.}

\maketitle

\section{Introduction}
\label{sec:introduction}

Last year, Hastings and Haah introduced a new quantum error correcting code, the honeycomb code, so-named due to its placement on a honeycomb lattice of physical qubits \cite{hastings2021dynamically}.
The logical qubits of the honeycomb code are defined by pairs of anti-commuting logical observables that change over time.  
This sidesteps a well-known issue with building a subsystem code out of Kitaev's honeycomb model, namely that it encodes no \emph{static} logical qubits \cite{kitaev2006anyons, suchara2011constructions, lee2017topological, wootton2021hexagonal}.

The honeycomb code's realization on a degree-3 (or less) grid of qubits has practical hardware merits \cite{chamberland2020topological}, and its construction out of weight-two parity measurements makes it natural for architectures where such operations are native (such as Majorana-based architectures \cite{chao2020optimization}).  
Historically, requiring this level of sparsity and locality has led to codes with steep losses in performance \cite{bacon2006operator, suchara2011constructions, li2019compass}.
But the honeycomb code's low-weight parity checks and succinct measurement cycle have yielded a robust memory under various circuit-level noise models \cite{gidney2021honeycombmemory}.
Altogether, the honeycomb code provides a foundation worth considering when building a fault-tolerant quantum computer.

We were originally attracted to the honeycomb code as a test case for Stim~\cite{gidney2021stim}, an open-source tool for simulating and analyzing stabilizer circuits and stabilizer codes.
Because the honeycomb code breaks implicit assumptions about how logical qubits are typically defined, it provided an excellent challenge to field test Stim's capabilities.
Using Stim, we numerically analyzed the performance of the honeycomb code in \cite{gidney2021honeycombmemory}.  
Our Monte Carlo simulations of the honeycomb code under circuit-level noise confirmed that it had a competitive threshold and low qubit overhead, particularly compared to other low-connectivity codes.

When we originally analyzed the honeycomb code, it lacked an elegant embedding onto a planar lattice.
We ended \cite{gidney2021honeycombmemory} noting that this was an important open problem, because embedding on a plane (instead of a torus) is a necessary ingredient for using the honeycomb code on two-dimensional architectures.
Fortunately, this problem was quickly solved by Haah and Hastings~\cite{haah2021boundaries}. 

In this paper, using much of the same software infrastructure from \cite{gidney2021honeycombmemory}, as well as some new capabilities, we quantify the performance of this new planar honeycomb code.
We perform Monte Carlo sampling on planar honeycomb patches, estimate their threshold, and project how many physical qubits they require to reach the teraquop regime (i.e. to hit one-in-a-trillion error rate logical operations).
In addition, we make a few modifications to the boundaries proposed by Haah and Hastings that may prove advantageous.
We use boundaries that have no extra connectivity requirements beyond those that naturally sit in the bulk of the honeycomb code.
A nice bonus is that these boundaries can be found easily, using a construction that cuts them out of a larger lattice.
Along the way, we point out important error mechanisms we missed while writing \cite{gidney2021honeycombmemory}, how we caught them, and leverage this information to optimize the patch shape that we use.

The paper will proceed as follows.  
In \sec{boundaries}, we define our version of boundaries for the planar honeycomb code. 
We also discuss different geometries of patches and their effective distances.
Then in \sec{sims}, we present simulation results that compare this planar honeycomb code to the periodic honeycomb code we benchmarked in \cite{gidney2021honeycombmemory}.
We then make our conclusions in \sec{conclusion}.
\app{noise} describes the noise models we used in the main text.
\app{extra_figures} includes some additional figures that were too detailed to fit into the main text.
\app{hardware_em3} describes some more physically-motivated noise models for direct parity measurements. 
\app{example_circuit} specifies the exact circuit of a small planar honeycomb code memory experiment.

\section{The Honeycomb Code with Boundaries}
\label{sec:boundaries}

\subsection{Defining the bulk}

The bulk of the honeycomb code is defined on a hexagonal lattice with physical qubits placed on the vertices.  
The faces of the lattice are three-colorable, and we associate to each color one of the three non-identity Pauli operators.  
In diagrams we associate $X$-type operators to red-colored faces, $Y$-type operators to green-colored faces, and $Z$-type operators to blue-colored faces.  
These faces will form the six-body stabilizers that we repeatedly measure to check for errors. 

Each edge of the hexagonal lattice travels between two faces of the same type, and is identified as being of that same type.
These edges will form the two-body measurements we perform - with Pauli type determined by color.
In each round we measure all edges of a particular color.
For example, an edge traveling between two green faces is a green edge and represents a two-body $Y$-basis parity measurement.
To avoid information overload in diagrams, we often don't explicitly color the edges and rely on the reader to infer the edge type from the surrounding faces.

Each stabilizer's value is equal to the product of the six edge measurements that encircle the stabilizer's face.
The three-coloring of the faces guarantees that the encircling edges are composed of edges of the two Pauli types other than the face's Pauli type.
The value of a face's stabilizer is recomputed whenever the previous two edge measurement layers match the two Pauli types of the face's encircling edges.

In the periodic honeycomb code the edge colors are measured in a consistent cycle - red, green, blue, red, green, blue, etc.
During this process, the logical operators must move in order to avoid anti-commuting with the next layer's edge measurements.  
This dance requires multiplying some of the previous layer's edge measurement results into the logical operator - choosing a new logical representative using the newly inserted weight-two stabilizers of the code state.
See \cite{gidney2021honeycombmemory} for a comprehensive summary.

\subsection{Introducing boundaries}

\begin{figure}[ht!]
    \centering
    \resizebox{\linewidth}{!}{
        \includegraphics{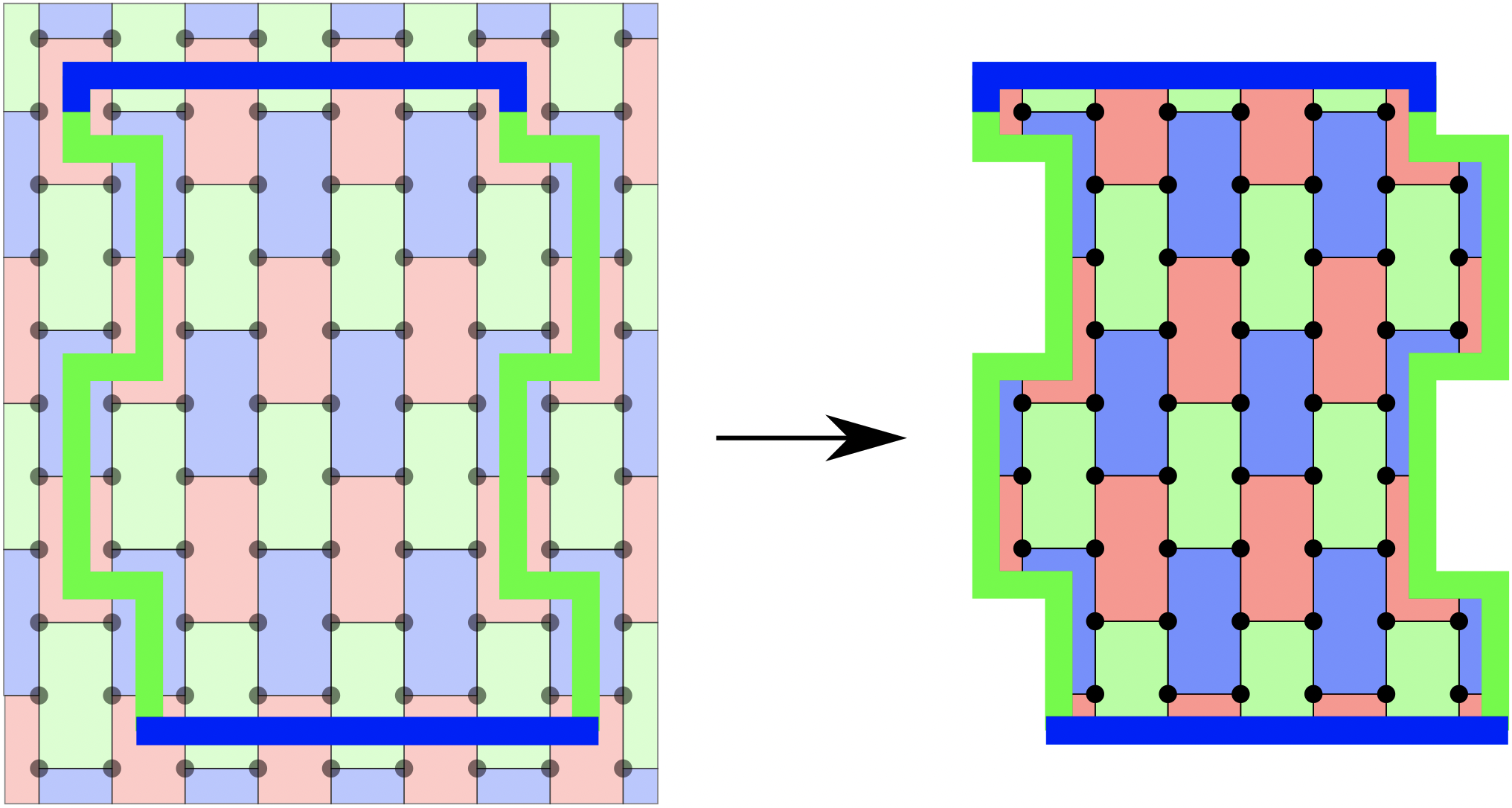}
    }
    \caption{
    Cutting a planar honeycomb code out of a hexagonal tiling.
    In the diagram, the cut line is blue when cutting blue edges and green when cutting green edges.
    A boundary of type A is formed by repeatedly cutting edges of type A.
    A logical qubit is formed when the boundaries are attached in an ABAB cycle.
    Boundary edges (i.e. edges that are cut) retain their color from the original lattice, but represent single-qubit measurements instead of two-qubit measurements.
    }
    \label{fig:cut_out_patch}
\end{figure}

To form boundaries of a particular color, we begin with an unbounded honeycomb code and cut out a finite sized patch - see \fig{cut_out_patch}.
We first cut through edges of one color $A$, then edges of a second color $B$, then edges of color $A$, then finally edges of color $B$ closing off the cut.
The edge color being cut through defines the type of the boundary along that part of the cut.
The $A$-$B$-$A$-$B$ pattern around the patch is analogous to the $X$-$Z$-$X$-$Z$ pattern of boundaries around surface code patches.

All qubits outside the cut are excluded.
The faces and edges inside the cut retain their roles as stabilizers and parity measurements.
This leads to face stabilizers of smaller sizes, and edge measurements along the boundary which correspond to single-qubit measurements.
Note that no extra qubit connectivity is required at the boundaries beyond what is available in the bulk.
This is a desirable property because it allows the underlying lattice of physical qubits to be homogeneous.
It doesn't restrict where logical qubits can be placed.

\begin{figure}[ht!]
    \centering
    \resizebox{\linewidth}{!}{
        \includegraphics{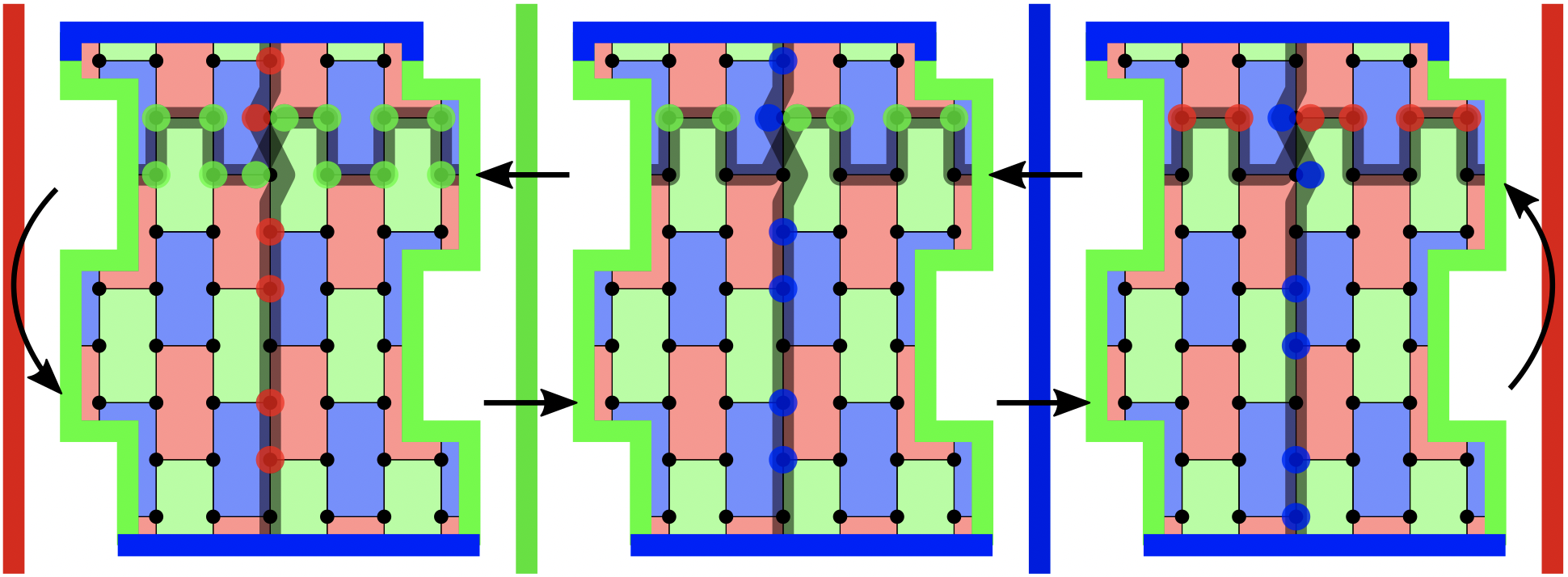}
    }
    \caption{
    Measurement ordering and observable locations in a planar honeycomb patch.
    The cycle of arrows indicates edge measurement layers alternating between X,Y,Z and X,Z,Y orderings.
    The colored vertical bars represent measuring all edges of the corresponding color (e.g. the green bar represents measuring all Y type edges).
    Fully crossing a vertical bar represents performing the corresponding edge measurements, and then multiplying the edges along an observable's path into that observable.  
    Note that red edges are never multiplied into the observable.
    The two anti-commuting observables are supported along the horizontal and vertical shaded paths, with colored circles indicating their Pauli product at that moment. 
    Note that the left-hand red measurement layer cannot be followed by a blue measurement layer, because the uppermost red operator in the vertical observable would anti-commute with a blue boundary edge measurement whether or not red edges had been multiplied into the observable.
    Similarly, the right-hand red measurement layer cannot be followed by a green measurement layer.
    }
    \label{fig:two_round_cycle}
\end{figure}

A logical qubit is formed out of two anti-commuting observables.
These observables must commute with the stabilizers of the honeycomb code.
Additionally, between each round of edge measurements, both observables must commute with the preceding edge measurements and also the upcoming edge measurements.
Unfortunately, if we reuse the cyclic $X$-$Y$-$Z$ edge measurement cycle that was used in the periodic honeycomb code, we will run into a problem where the logical operators unavoidably anti-commute with some of the single-qubit measurements at the boundaries.
However, it turns out that this problem can be avoided by alternating between two edge measurement orderings~\cite{hastings2021dynamically}, such as $X$-$Y$-$Z$ and $X$-$Z$-$Y$.
The resulting scheduling successfully extracts the required stabilizers while allowing the logical observables to dance around the edge measurements - see \fig{two_round_cycle}.

\begin{figure}[ht!]
    \centering
    \resizebox{\linewidth}{!}{
        \includegraphics{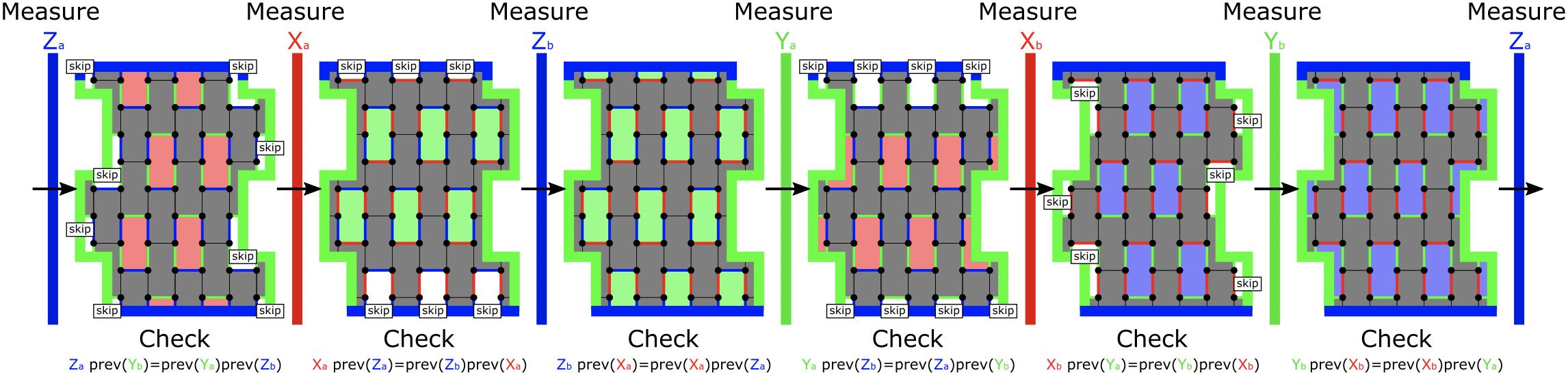}
    }
    \caption{
    Face stabilizers collected after each edge measurement layer.  
    Each edge measurement layer has a subscript $a$ or $b$ to distinguish its location within the two-round X,Y,Z-then-X,Z,Y cycle of edge measurements. 
    The faces collected after a particular layer are color-filled.  
    Boundary faces of the correct color but which could \emph{not} be collected in this layer due to commutation constraints - see \fig{detectors} - are white with `skip' labels.  
    In particular, each boundary face stabilizer can only be collected in a layer that measures all of the face's boundary edges.
    Consequently, the red faces in the corners, having boundary edges of different colors, are \emph{never} collected.  
    The remaining different-color faces are shaded grey.  
    The detectors themselves - namely the measurement comparisons to be checked - are labeled by `Check'.
    Each check is evaluated on the edges surrounding each of the colored-in faces at that step.
    }
    \label{fig:parity_check_cycle}
\end{figure}

\begin{figure}[ht!]
    \centering
    \resizebox{0.85\linewidth}{!}{
        \includegraphics{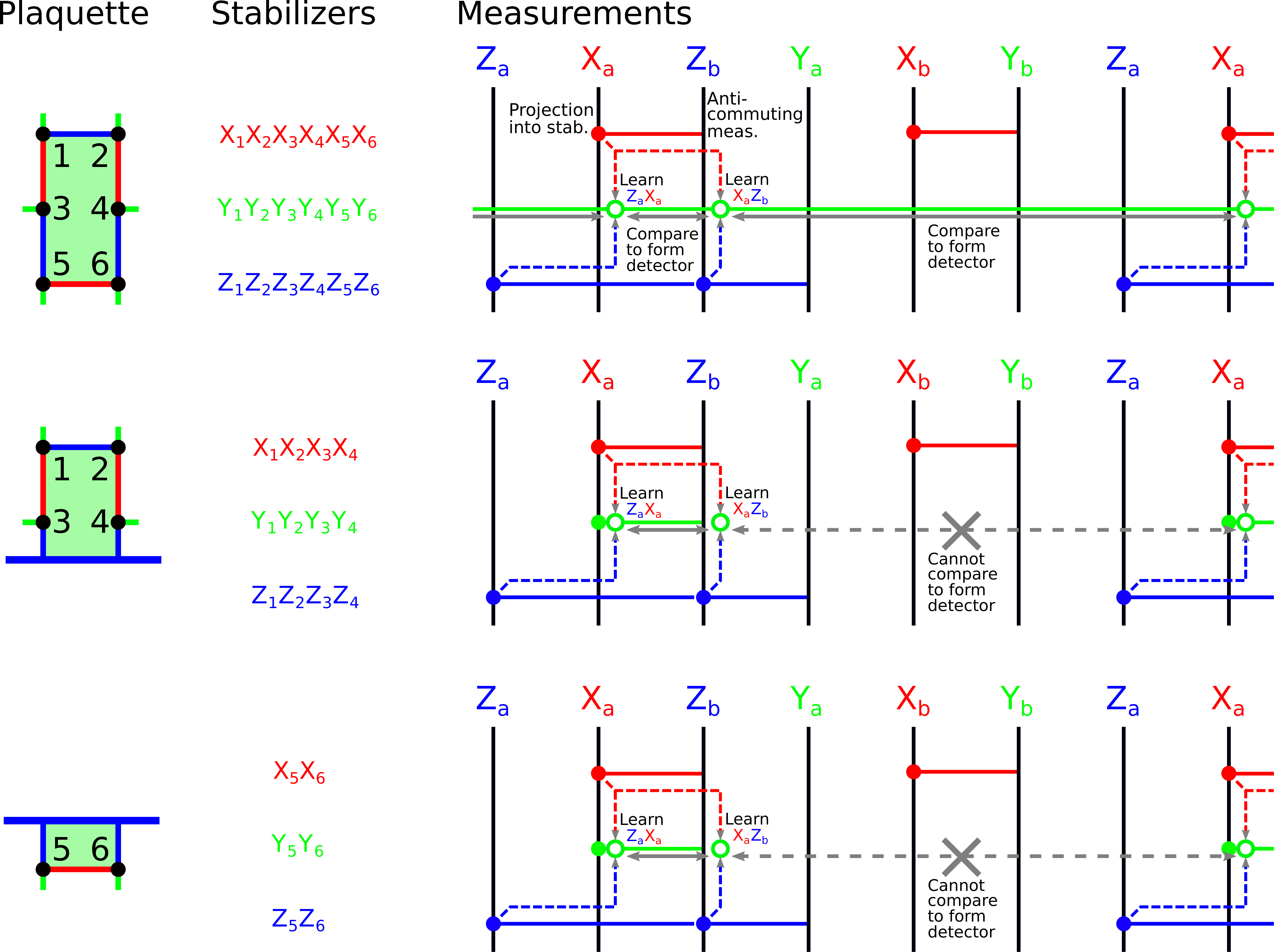}
    }
    \caption{
    The evolution of stabilizers for a bulk plaquette, for which two detectors are formed every six edge measurement rounds, and for both kinds of a boundary plaquette, for which one detector is formed every six edge measurement rounds.  The vertical black bars represent the edge measurements, while the horizontal colored lines represent the time during which a given stabilizer is valid (i.e. stabilizes the state of the qubits on the plaquette).  This illustrates that bulk plaquettes feature a stabilizer that remains consistently valid through the measurement cycle. Filled circles represent measurements that project the plaquette qubits into the subspace of the given stabilizer.  Open circles represent occasions where we learn a stabilizer by combining other measurements, as indicated by dashed arrows. Grey arrows indicate detectors, which are comparisons of consistent values of stabilizers (or equivalently, products of all the check measurements from which we learned the stabilizers in question). These detectors should yield deterministic parity in the absence of noise, and detect errors in the presence of noise.
    }
    \label{fig:detectors}
\end{figure}

A potential drawback of alternating between orderings is that there are longer gaps of time between measuring certain stabilizers, resulting in some stabilizers accumulating more noise than others.
These manifest as zig-zag patterns in the rate that detection events occur as time passes - see \fig{detection_fractions_over_time}.
Another complication is that, due to destabilizing single-qubit boundary edge measurements, certain boundary detectors (i.e. deterministic comparisons of face stabilizer outcomes) are collected only once per six edge measurements, compared to twice per six edge measurements in the bulk - see \fig{parity_check_cycle} and \fig{detectors}.
We were initially worried that these issues would hurt performance, compared to the periodic honeycomb code.
Fortunately, as we will show, these changes do \emph{not} qualitatively affect the robustness of the code.

\subsection{Choosing the patch}
\label{sec:patches}

One of the benefits of defining the planar honeycomb code in terms of cutting a patch out of the unbounded honeycomb code is \emph{simplicity}.  
It's not hard to figure out how to move in a particular direction while cutting a specific edge type.
Additionally, every qubit and operation in the planar code is a qubit or operation inherited from the unbounded honeycomb code.
From a circuit perspective, the cutting approach is purely subtractive.
It removes operations, or shrinks operations, but never introduces operations.
This means it's not hard to figure out how to ``cap off'' the boundaries (i.e. which operations need to occur there in order for the circuit to be correct).
This is analogous to the boundaries of the surface code, which can be introduced in a similar way.

It is worth noting that not every edge measurement we make is used to form a detector.  
In \fig{parity_check_cycle}, immediately after the edge measurement step labeled $X_a$, consider a red edge measurement incident to a green face along the top boundary. This case is also illustrated at the bottom of \fig{detectors}.
This red edge measurement can only contribute to the blue stabilizer below it and the green stabilizer above it.
The measurement doesn't contribute to the blue stabilizer because the step $X_a$ is sandwiched between two blue edge steps, and blue stabilizers are only measured when a red edge measurement step is next to a green edge measurement step.
The measurement does appear in the equation used for checking the green stabilizer, but it appears \emph{twice} (once on the left and once on the right).
These two contributions cancel out and can be omitted.
Therefore this specific edge measurement is never used.
It could be removed from the circuit, freeing up spacetime that could be used for other purposes.
That would likely be beneficial but, to keep the definition of our boundaries simple and uniform, we leave things as they are instead of removing the unused measurements.

In \cite{gidney2021honeycombmemory}, we used honeycomb codes covering regions of data qubits with a width:height aspect ratio of 2:3.
We picked this ratio by carefully thinking about what the fastest-horizontally-moving and fastest-vertically-moving error mechanisms would be.
For this paper, we used a new feature in Stim which can automatically search for the smallest undetectable set of ``graphlike'' errors that cause a logical error.
A graphlike error is a physical error that produces at most two detection events.
For example, a single qubit X error on a data qubit immediately after a measurement is a graphlike error, as it will trigger the detectors associated with the two non-commuting incident faces. An example of a non-graphlike error in the honeycomb code is a single measurement error, which causes 4 detection events - two for each of the faces it is incident to.  
The graphlike code distance may be larger than the true code distance, because it neglects the possibility that errors that trigger multiple detectors may produce faster moving errors. However, it can be computed cheaply by reduction to a shortest path problem. The graphlike error distance therefore is a useful heuristic for choosing e.g. patch dimensions, but the Monte Carlo sampling is the real arbiter of code performance, as it will include the influence of all elements of the error model. Indeed, departures from error scaling matching the graphlike code distance would be a good indication that non-graphlike errors are involved in the dominating error mechanism.

\begin{figure}[ht!]
    \centering
    \resizebox{\linewidth}{!}{
        \includegraphics{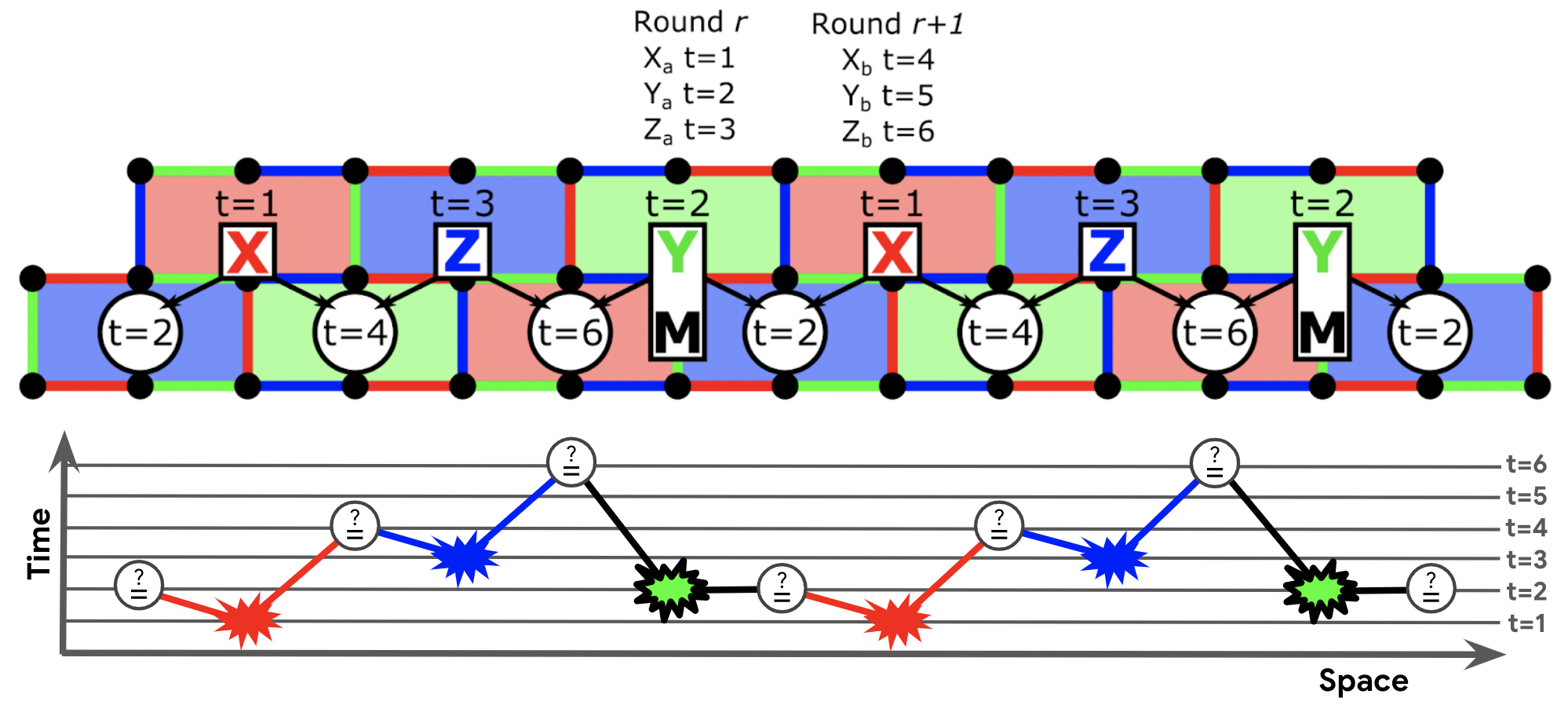}
    }
    \caption{
    A sparse long distance error mechanism along a vertical line of qubits in the honeycomb code, which was found by a search for short undetectable sets of graphlike errors.
    Each white box represents a single physical error, with the kind of error indicated by the text inside the box.
    The `YM` error corresponds to both a Y error on one qubit and a classical flip of the reported measurement outcome for the indicated edge measurement.
    Each white circle represents a detector from the circuit.
    The number in the circle is the edge measurement layer where that detector is checked, potentially producing a detection event.
    An arrow from a white box to a white circle indicates that the error triggers the detector to produce a detection event.
    Triggering a detector twice cancels out the detection event.  
    Below, we plot the error string traversing the lattice through time as well as space, illustrating the importance of the `YM` measurement in resolving the time ordering.
    }
    \label{fig:bad_error}
\end{figure}

The graphlike code distance search immediately found error mechanisms that moved faster vertically than we thought was possible in the SD6 and SI1000 circuits. We show an example of such an error mechanism in \fig{bad_error}, where a non-obvious pattern of `X`, `Z` and `YM` on every 2nd qubit along a vertical line connect the boundaries without producing any detection events. A `YM` error is a combination of a single qubit `Y` error and a classical measurement error during one edge measurement. In the EM3 error model, the `YM` error is a native error case. In SD6 and SI1000 circuits, the `YM` error is a hook error caused by inserting an physical error on the measurement ancilla halfway through the edge measurement. 
Note that you can't simply replace the `YM' hook error with a `Y' data error, due to time ordering constraints.

With hindsight on our side, we looked more closely at the data we collected for \cite{gidney2021honeycombmemory} and realized this is actually noticeable from the plots included in the appendix of the paper (which separately showed the logical error rates for the horizontal and vertical observables).
Consequently, in this paper, for the SD6 and SI1000 layouts, we now use a 1:2 aspect ratio instead of a 2:3 aspect ratio.

The graphlike code distance search also helped us choose a more efficient lattice geometry.
Our initial implementation of the planar honeycomb code used what we now call a sheared layout - see \fig{sheared}.
We expected this layout to perform better because it made all the boundaries identical, up to rotations of the hexagonal tiling and permutations of the colors.
The graphlike code distance search revealed that this expectation was wrong.
\tbl{graphlike_distances} includes the horizontal and vertical graphlike code distances for the three error models and both straight and sheared boundaries, demonstrating the lower horizontal graphlike code distance of the sheared patch.
The issue is that, because of the rotational symmetry in a hexagonal tiling, there are fast moving diagonal error mechanisms analogous to the fast moving vertical error mechanisms that forced us to change the patch aspect ratio.
Shearing the layout preserves the horizontal distance between the two vertical boundaries, but it reduces the diagonal distance.
Based on this information, we switched to using vertical boundaries that moved straight up-and-down - see \fig{cut_out_patch} - squiggling back and forth in a graphlike-code-distance-maximizing way.

\begin{table}[ht!]
    \centering
    \resizebox{\linewidth}{!}{
    \begin{tabular}{| r | c | c | c | c | c | c | c | c | c | c | c | c | c |}
    \hline
    width >> height &            $h=3$ &            $h=6$ &            $h=9$ &           $h=12$ &           $h=15$ &           $h=18$ &           $h=21$ &           $h=24$ &           $h=27$ &           $h=30$ &           $h=33$ &           $h=36$ &           $h=39$ \\
    \hline
                 EM3 &                1 &                2 &                3 &                4 &                5 &                6 &                7 &                8 &                9 &               10 &               11 &               12 &               13 \\
                 SD6 &                1 &                3 &                4 &                6 &                7 &                9 &               10 &               12 &               13 &               15 &               16 &               18 &               19 \\
              SI1000 &                1 &                3 &                4 &                6 &                7 &                9 &               10 &               12 &               13 &               15 &               16 &               18 &               19 \\
          (sheared) EM3 &                1 &                2 &                3 &                4 &                5 &                6 &                7 &                8 &                9 &               10 &               11 &               12 &               13 \\
          (sheared) SD6 &                1 &                3 &                4 &                6 &                7 &                9 &               10 &               12 &               13 &               15 &               16 &               18 &               19 \\
       (sheared) SI1000 &                1 &                3 &                4 &                6 &                7 &                9 &               10 &               12 &               13 &               15 &               16 &               18 &               19 \\
    \hline
    height >> width &            $w=2$ &            $w=3$ &            $w=4$ &            $w=5$ &            $w=6$ &            $w=7$ &            $w=8$ &            $w=9$ &           $w=10$ &           $w=11$ &           $w=12$ &           $w=13$ &           $w=14$ \\
    \hline
                 EM3 &                1 &                1 &                2 &                2 &                3 &                3 &                4 &                4 &                5 &                5 &                6 &                6 &                7 \\
                 SD6 &                1 &                2 &                3 &                4 &                5 &                6 &                7 &                8 &                9 &               10 &               11 &               12 &               13 \\
              SI1000 &                1 &                2 &                3 &                4 &                5 &                6 &                7 &                8 &                9 &               10 &               11 &               12 &               13 \\
          (sheared) EM3 &                1 &              N/A &                2 &              N/A &                3 &              N/A &                4 &              N/A &                5 &              N/A &                6 &              N/A &                7 \\
          (sheared) SD6 &                1 &              N/A &                3 &              N/A &                4 &              N/A &                6 &              N/A &                7 &              N/A &                9 &              N/A &               10 \\
       (sheared) SI1000 &                1 &              N/A &                3 &              N/A &                4 &              N/A &                6 &              N/A &                7 &              N/A &                9 &              N/A &               10 \\
    \hline
\end{tabular}
    }
    \caption{
    Horizontal and vertical graphlike code distances of planar honeycomb patches under circuit noise.
    Computed in four minutes using stim.Circuit.shortest\_graphlike\_error.
    }
    \label{tbl:graphlike_distances}
\end{table}

\section{Simulations}
\label{sec:sims}

In this section, we use Monte Carlo sampling to quantify the performance of the planar honeycomb code.
We compute the threshold, lambda factors, and teraquop footprints (the number of qubits required to hit one-in-a-trillion failure rates) and to compare these numbers to the ones we determined for the periodic honeycomb code in \cite{gidney2021honeycombmemory}.

Our methodology is based on memory experiments simulated using Stim~\cite{gidney2021stim} and decoded using an internal software tool that performs uncorrelated and correlated minimum weight perfect matching.
A memory experiment initializes a logical qubit into a known state, idles the logical qubit for some number of rounds while performing error correction to protect it against noise, and then measures whether or not the logical qubit is still in the intended state.
The Python code used to generate circuits, run simulations, and render plots is attached to this paper in the ancillary file directory \path{code/} and is also available online at \href{https://github.com/strilanc/honeycomb_threshold}{github.com/strilanc/honeycomb\_threshold}.

We worked with three different variations of the memory experiment: H-type, V-type, and EPR-type.
All three variations use the same physical operations to preserve their logical state, but differ in which state they prepare and measure.
The H-type experiment uses noisy transversal initialization/measurement to fault-tolerantly prepare and measure the horizontal observable.
The V-type experiment uses noisy transversal initialization/measurement to fault-tolerantly prepare and measure the vertical observable.
The EPR-type experiment uses magically noiseless operations to prepare a perfect EPR pair between the logical qubit and a magically noiseless ancilla qubit.
The Bell basis measurement at the end, between the logical qubit and the ancilla qubit, also uses magically noiseless operations.

While we do not report Monte Carlo simulations of the EPR-type experiment, we used it extensively when writing and debugging our code.
The EPR experiment checks that the honeycomb code circuit preserves an entangled state, verifying that both observables are simultaneously protected and proving the logical qubit is actually a qubit.
We didn't use the EPR type experiment for benchmarking because we were worried about distortions from the magically noiseless time boundaries.
Also, we prefer to report benchmarks that don't impose requirements that a real quantum computer couldn't meet.

In plots, unless otherwise specified, the error rate that we focus on is ``combined code cell logical error rate".
By ``code cell" we mean that the error rates refer to the error rate per block of $d$ rounds where $d$ is the graphlike code distance (the number of graphlike physical errors required to produce an undetectable logical error).
This differs from other per-round error rates often reported in the literature, but is more relevant to the error rate of a logical operation, which we expect to have a timelike extent on the order of the code distance.  
Beware that our definition couples the number of rounds in a code cell to the noise model, because different noise models have different graphlike code distances.
These differences are relatively small on all log-log plots but good to keep in mind.
Also note that our memory experiment circuits are $3d$ rounds long.

The ``combined'' in ``combined code cell logical error rate'' refers to the fact that we are extrapolating the logical error rate for an arbitrary state by combining the error rates from the H-type and V-type experiments (which use specific states not vulnerable to all errors).
We combine the error rates by assuming they are independent errors and that both must not occur.
If $E_H$ and $E_V$ are the H-type and V-type error rates then we define the combined error rate $E_{HV}$ to be

\begin{equation}
    E_{HV} = 1 - (1 - E_{\text{H}}) (1 -  E_{\text{V}})
    \label{eq:combined_logical_error}
\end{equation}

We use three error models in this paper (see \tbl{models}).
The first error model we use is SD6 (standard depolarizing 6-step cycle), a standard circuit model with homogeneous component errors relevant to the QEC literature.
The second is SI1000 (superconducting-inspired 1000 $ns$ cycle), a superconducting-inspired error model where, for example, the measurements are significantly noisier than the gates.  
Finally, we adopt the Majorana-inspired EM3 model (entangling-measurement 3-step cycle), which includes two-qubit entangling measurements, to compare with the surface code benchmarking performed in \cite{chao2020optimization}. Variations on this EM3 error model are discussed in \app{hardware_em3}.

\begin{figure}[ht!]
    \centering
    \resizebox{\linewidth}{!}{
        \includegraphics{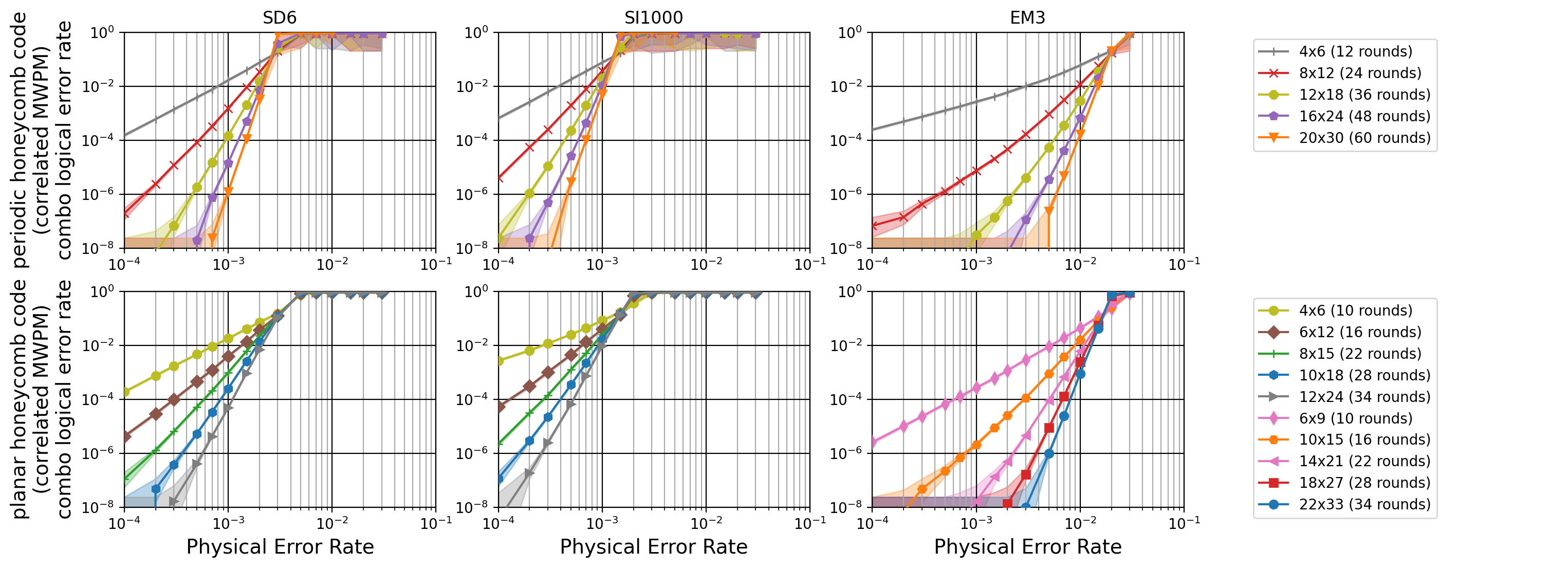}
    }
    \caption{
    Logical error rate plots of the performance of a periodic and planar honeycomb memory.
    Note that, due to our optimizations of the patch shape, the old periodic honeycomb data (from \cite{gidney2021honeycombmemory}) uses different patch sizes.
    Also, the new planar honeycomb data uses different patch sizes for different error models.
    The plotted logical error rate is the combined code cell logical error rate (see \eq{combined_logical_error}).
    Color highlights indicate statistical uncertainty from Monte Carlo sampling.
    The highlight covers hypothetical logical error rates with a Bayes factor of at most 1000 vs the maximum likelihood hypothesis probability, assuming a binomial distribution.
    In both the planar and periodic cases, we see the noticeable effect of damaging correlated errors produced by the EM3 noise model, which result in non-linear behaviour at low error rates, coinciding with \cite{gidney2021honeycombmemory}.
    The raw planar honeycomb logical error rate data we collected for this paper is in ancillary file ``stats\_planar\_honeycomb.csv".
    The periodic honeycomb logical error rate data we are comparing against is in ancillary file ``stats\_from\_previous\_paper.csv" (copied over from \cite{gidney2021honeycombmemory}).
    }
    \label{fig:thresholds}
\end{figure}

\begin{figure}[ht!]
    \centering
    \resizebox{\linewidth}{!}{
        \includegraphics{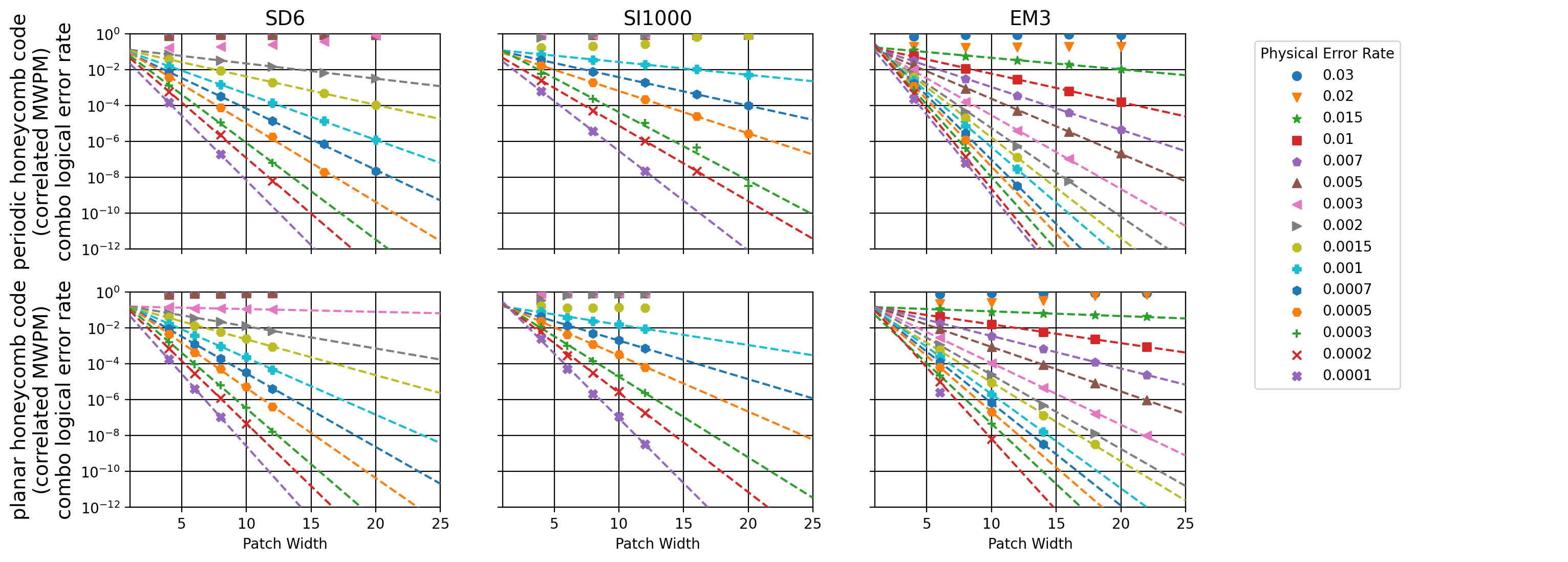}
    }
    \caption{
    Line fit plots.
    These are used to project the patch size required to achieve a desired logical error rate per logical operation for various physical error rates, assuming error suppression grows linearly in log space versus patch width.
    Fits are only included for error rates which demonstrate error suppression.  From this, we can surmise a threshold within $0.2\%-0.3\%$ for SD6, a threshold within $0.1\%-0.15\%$ for SI1000, and a threshold within $1.5\%-2.0\% $ for EM3.  
    The corresponding thresholds in the surface code are $0.5\%-0.7\%$ for SD6~\cite{gidney2021honeycombmemory}, $0.3\%-0.5\%$ for SI1000~\cite{gidney2021honeycombmemory}, and at least $0.2\%$ for EM3~\cite{chao2020optimization}.
    The performance reversal in the Majorana-inspired EM3 model is due to the large number of ancilla and extra operations required to extract syndrome information in the surface code compared with the honeycomb code.
    }
    \label{fig:line_fits}
\end{figure}

\fig{thresholds} contains the raw logical error rates for both the periodic and planar honeycomb codes, while \fig{line_fits} shows line fits projecting logical error rates at larger code distances.
Both are computed using a custom MWPM decoder that includes correlations between the two connected components of the error graph \cite{gidney2021honeycombmemory, fowler2013optimal}. 
From that data, we conclude that the threshold of the planar honeycomb code is not appreciably lower than the periodic honeycomb code.  
Note that there was reason to believe the threshold - which is often set by the bulk behavior - might decrease due to the modification of the bulk syndrome cycles.
In the SD6 model it even appears higher, likely an artifact of scaling with more carefully chosen aspect ratios.

Comparing with the surface code thresholds obtained in an identical error model in \cite{gidney2021honeycombmemory}, the threshold of the honeycomb code is approximately half that of the surface code in SD6 and SI1000.
Similar to \cite{gidney2021honeycombmemory}, in the EM3 model using two-qubit entangling measurements, the threshold is between 5x-10x greater than the surface code in a similar model \cite{chao2020optimization}.

\begin{figure}[ht!]
    \centering
    \resizebox{\linewidth}{!}{
        \includegraphics{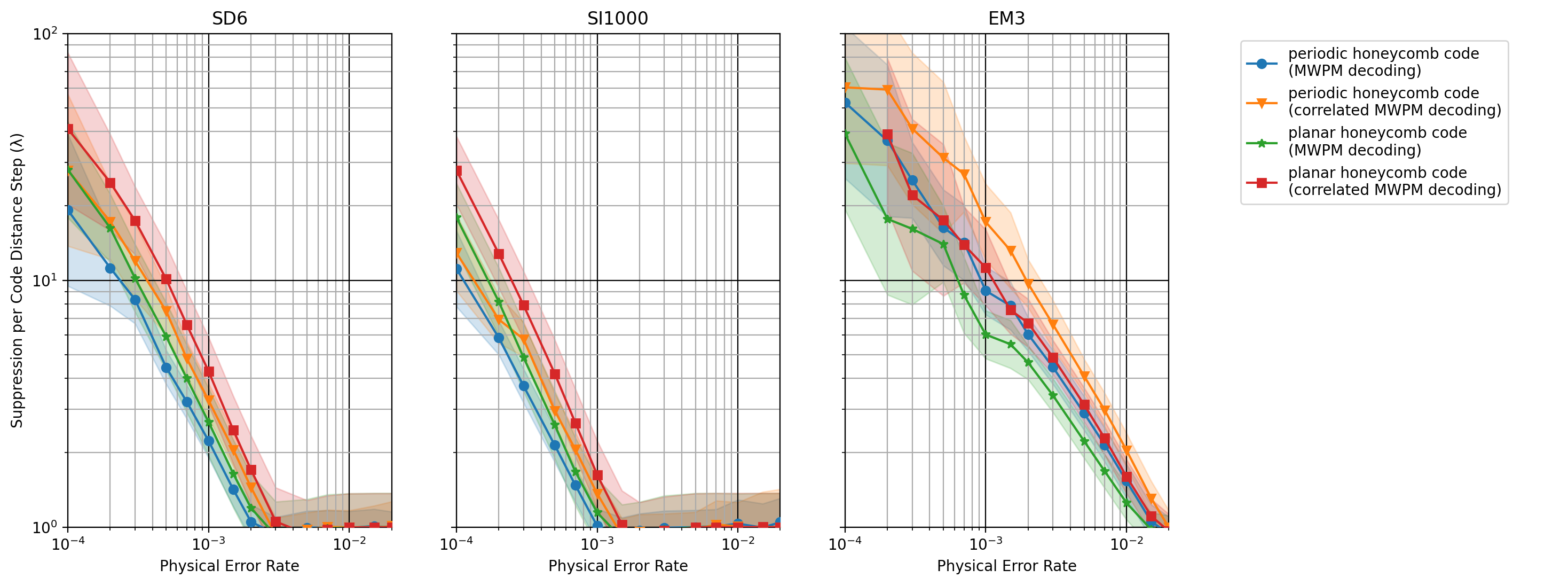}
    }
    \caption{
    Lambda factor plots.
    These project the multiplicative factor by which the logical error rate is reduced each time the code distance is increased by 2 at different noise rates.
    Estimated using the slope of the line fits from \fig{line_fits}.
    The color highlighting indicates how far each point can be moved by using line fits whose least-squares error terms are at most $1.0$ higher than the error term for the optimal least-squares fit when fitting against the natural log of the logical error rates.
    }
    \label{fig:lambdas}
\end{figure}

\begin{figure}[htb!]
    \centering
    \resizebox{\linewidth}{!}{
        \includegraphics{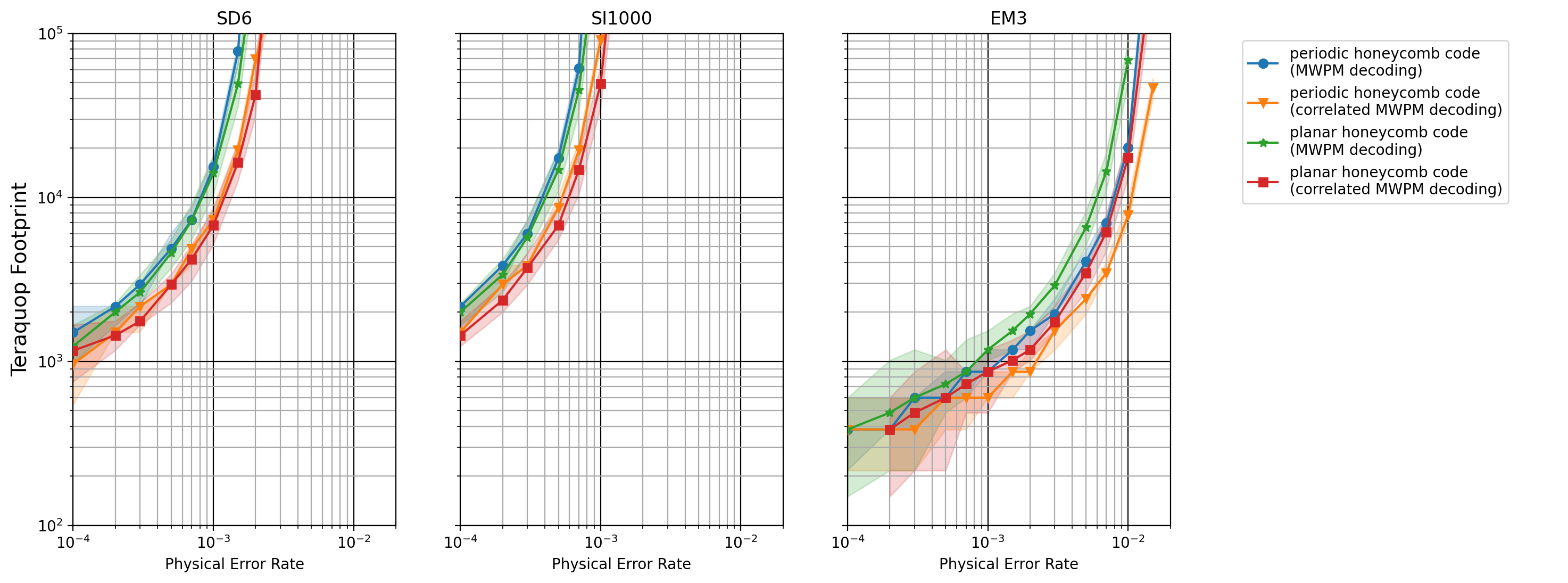}
    }
    \caption{
    Teraquop footprint plots.
    This projects the number of physical qubits required to store a logical qubit capable of performing approximately a trillion logical operations - assuming each takes approximately distance-$d$ rounds to execute.
    The code distance required is projected using the line fits from \fig{line_fits}.
    The projected code distance is rounded up to an achievable set of code parameters, and then converted to a qubit count.
    The color highlighting indicates how far each point can be moved by using line fits whose least-squares error terms are at most $1.0$ higher than the error term for the optimal least-squares fit when fitting against the natural log of the logical error rates.
    }
    \label{fig:teraquop_footprints}
\end{figure}

\fig{lambdas} and \fig{teraquop_footprints} plot the lambda factors and teraquop footprints, respectively.  
From \fig{lambdas} we verify that lambda increases approximately linearly with error rate well below threshold.
Ultimately, the teraquop footprint plot answers the most important question of how much qubit overhead is required to hit a target error rate.
Essentially, we observe that enforcing planar constraints incurs no qualitative cost to the qubit overheads required for a teraquop memory.
At a physical error rate of $10^{-3}$, a planar teraquop honeycomb memory costs approximately 7000/50000/900 qubits in the SD6/SI1000/EM3 models.
SI1000 has a particularly bad teraquop footprint at this physical error rate because $10^{-3}$ is close to threshold in the SI1000 model.
Correspondingly, the SI1000 teraquop footprint drops dramatically as the physical error rate decreases below $10^{-3}$ (see \fig{teraquop_footprints}).

\section{Conclusion}
\label{sec:conclusion}

The honeycomb code is a promising candidate for a sparsely connected quantum memory.  
While it is not as robust as the surface code in the more standard SD6 and SI1000 error models, it boasts a teraquop footprint within an order of magnitude of the surface code at physically plausible error rates.  
In the EM3 error model, it significantly outperforms previous studies of the surface code \cite{chao2020optimization}.

Unlike the planar variant of the surface code, where the circuit is nearly identical to the toric code, the planar variant of the honeycomb requires an appreciable departure from its periodic cousin in terms of the ordering of operations.
In spite of these changes, we observe that the planar honeycomb code performs essentially as well as the periodic variant studied in \cite{gidney2021honeycombmemory}.

That being said, we caution the reader that our results are \emph{not} a direct comparison of the planar and periodic honeycomb codes.
We made multiple changes that affect performance, and are comparing the planar honeycomb code after these changes to the periodic honeycomb code before these changes.
The most significant change is that we switched from using the patch width as the code distance to using the graphlike code distance as the code distance.
For example, because the target code distance of EM3 circuits was effectively cut in half by this change, and because the code distance determines how many rounds of idling we count as ``one logical operation", it's as if we were previously requiring the periodic EM3 honeycomb code to achieve a logical error rate of one per \emph{two} trillion operations when computing its teraquop footprint.

Ultimately, our results cement the planar honeycomb code as a promising candidate for building a scalable fault-tolerant quantum computer.
We hope that the smallest conceivable distance-3 instantiation of the code, using a $4\times6$ patch of data qubits, can be tested experimentally in the future.

\section{Contributions}

All authors worked together extensively to understand the planar honeycomb code and to present it in a clear way.
Craig Gidney wrote the circuit generation, simulation, and plotting code.
Michael Newman did the majority of the paper writing.
Matt McEwen contributed refactored variants of the EM3 error model inspired by hardware, and revised the paper for publication.

\section{Acknowledgements}

We thank Austin Fowler for writing the internal software tool we used for syndrome decoding.
We thank Adam Paetznick, Christina Knapp, Nicolas Delfosse, Bela Bauer, Jeongwan Haah, Matthew B. Hastings, and Marcus P. da Silva for coordinating to make their independent results available to the public simultaneously with ours.
We thank Hartmut Neven for creating an environment where this research was possible.

\bibliographystyle{plainnat}
\bibliography{refs}

\appendix
\section{Error Models and Syndrome Cycles}
\label{app:noise}

We used essentially the same noise models that we used in \cite{gidney2021honeycombmemory}.
See \tbl{noise} and \tbl{models}.
There are two notable differences.

First, the definition of single qubit measurement error in \cite{gidney2021honeycombmemory} implicitly assumed the measurement was a demolition measurement.
The EM3 error model in this paper uses non-demolition single qubit measurements at the boundaries.
We introduced a ``truncated parity measurement" $M_{PI}(p)$ to account for this.

Second, in \cite{gidney2021honeycombmemory} we accidentally said that we were applying errors \emph{after} $M_{PP}$ measurements.
Actually, the code applies errors \emph{before} the measurement.

\begin{table}[h]
    \centering
    \resizebox{\linewidth}{!}{
    \begin{tabular}{|c|l|}
         \hline
         \textbf{Noisy Gate} & \textbf{Definition} \\
         \hline
         $\text{AnyClifford}_2(p)$ & \text{Any two-qubit Clifford gate, followed by a two-qubit depolarizing channel of strength $p$.} \\
         \hline
         $\text{AnyClifford}_1(p)$ & Any one-qubit Clifford gate, followed by a one-qubit depolarizing channel of strength $p$. \\
         \hline
         $\text{Init}_Z(p)$ & Initialize the qubit as $\ket{0}$, followed by a bitflip channel of strength $p$. \\
         \hline
         $M_Z(p)$ & Precede with a bitflip channel of strength $p$, and measure the qubit in the $Z$-basis. \\
         \hline
          & Measure a Pauli product $PP$ on a pair of qubits and, with probability $p$, choose an error
          \\ $M_{PP}(p)$ & uniformly from the set $\{I,X,Y,Z\}^{\otimes 2} \times \{\text{flip, no flip}\}$.
          \\&The flip error component causes the wrong result to be reported.
          \\&The Pauli error components are applied to the target qubits before the measurement.
          \\&If one of the Pauli operators is $I$, so that it is a single-qubit measurement (on the boundary), 
          \\&choose an error uniformly from the set $\{I,X,Y,Z\} \times \{\text{flip, no flip}\}$.\\ 
         \hline
         $\text{Idle}(p)$ & If the qubit is not used in this time step, apply a one-qubit depolarizing channel of strength $p$. \\
         \hline
         $\text{ResonatorIdle}(p)$ & If the qubit is not measured or reset in a time step during which other qubits are \\ &  being measured or reset, apply a one-qubit depolarizing channel of strength $p$. \\
         \hline
    \end{tabular}
    }
    \caption{
        Modified from \cite{gidney2021honeycombmemory}.
        Noise channels and the rules used to apply them.
        Noisy rules stack with each other - for example, Idle($p$) and ResonatorIdle($p$) can both apply depolarizing channels in the same time step.
        Note that our direct measurement error model is still using circuit noise (it is not a phenomenological noise model) - correlated errors are applied according to the support of each two-qubit measurement, coinciding with the error model in \cite{chao2020optimization}. For further discussion of this error model, see \app{hardware_em3}.
    }
    \label{tbl:noise}
\end{table}

\begin{table}[h]
    \centering
    \begin{tabular}{|r|l|l|l|}
        \hline
        \textbf{Abbreviation}
        &SD6
        &SI1000
        &EM3
        \\\hline
        \textbf{Name}
            & \begin{tabular}{@{}l@{}}Standard\\Depolarizing\end{tabular}
            & \begin{tabular}{@{}l@{}}Superconducting\\Inspired\end{tabular}
            & \begin{tabular}{@{}l@{}}Entangling\\Measurements\end{tabular}
        \\\hline
        \textbf{Noisy Gateset}
            &\noindent\begin{tabular}{@{}l@{}}
                $\text{CX}(p)$\\
                $\text{AnyClifford}_1(p)$\\
                $\text{Init}_Z(p)$\\
                $M_Z(p)$\\
                $\text{Idle}(p)$\\
            \end{tabular}
            &\begin{tabular}{@{}l@{}}
                \vspace{-0.25cm}
                {} \\
                $\text{CZ}(p)$\\
                $\text{AnyClifford}_1(p/10)$\\
                $\text{Init}_Z(2p)$\\
                $M_Z(5p)$\\
                $\text{Idle}(p/10)$\\
                $\text{ResonatorIdle}(2p)$\\
            \end{tabular}
            &\begin{tabular}{@{}l@{}}
                $\text{M}_{PP}(p)$\\
                $\text{M}_{PI}(p)$\\
                $\text{Init}_Z(p/2)$\\
                $\text{M}_Z(p/2)$\\
                $\text{Idle}(p)$\\
            \end{tabular}
        \\\hline
        \begin{tabular}{@{}r@{}}\textbf{Measurement}\\\textbf{Ancillae}\end{tabular}
            &Yes
            &Yes
            &No
        \\\hline
        \begin{tabular}{@{}r@{}}\textbf{Honeycomb}\\\textbf{Cycle Length}\end{tabular}
            & 6 time steps
            & 9 time steps ($\approx1000$ns)
            & 3 time steps
        \\\hline
    \end{tabular}
    \caption{
        Modified from \cite{gidney2021honeycombmemory}.
        The three noisy gate sets investigated in this paper.
        See \tbl{noise} for the exact definitions of the noisy gates.
        The superconducting-inspired acronym refers to an expected cycle time of about $1000$ nanoseconds \cite{chen2021exponential}.
    }
    \label{tbl:models}
\end{table}

\clearpage
\section{Supplementary Figures}
\label{app:extra_figures}

\begin{figure}[h]
    \centering
    \resizebox{\linewidth}{!}{
        \includegraphics{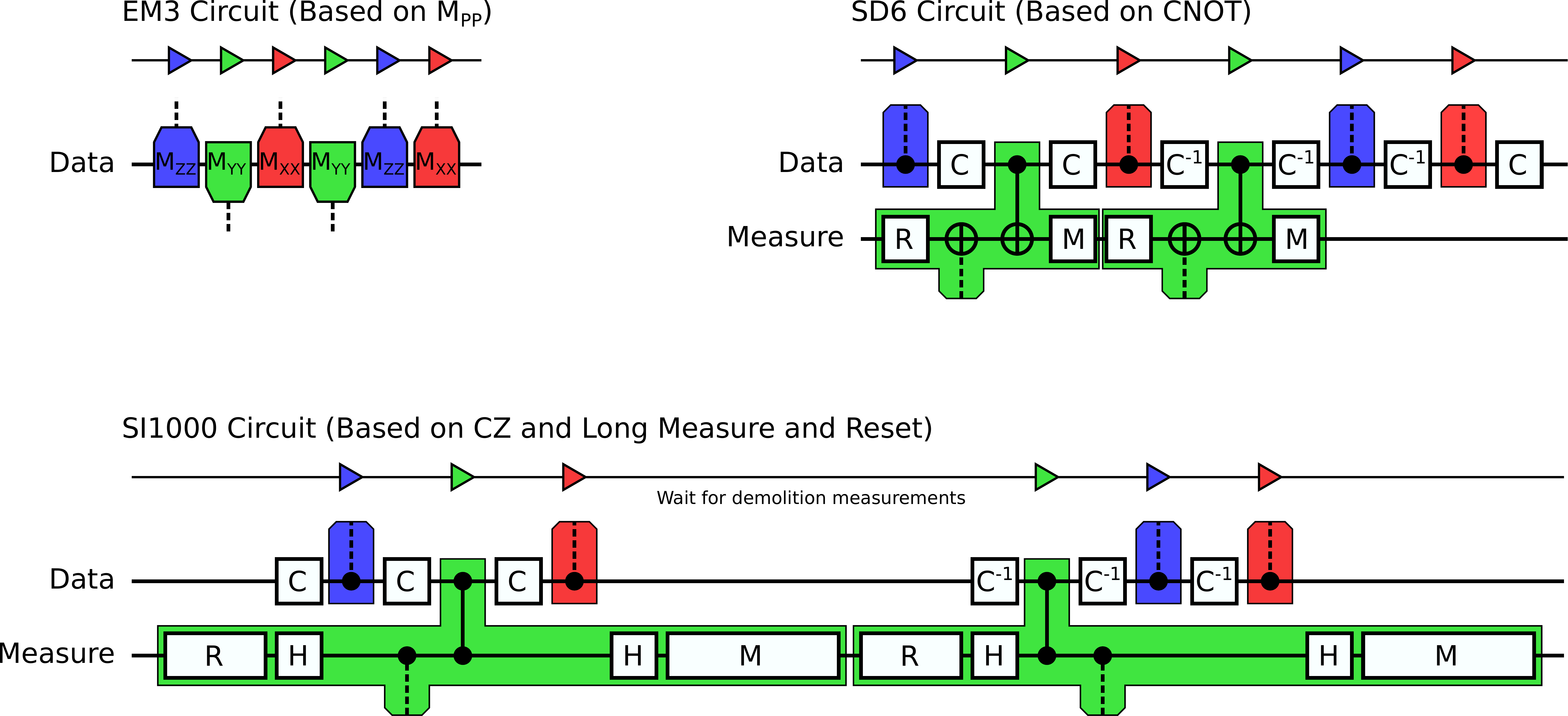}
    }
    \caption{
    Circuit cycles used for each of the three noisy gate sets considered in this paper.
    The entangling measurement noisy gate set (EM3) has the desired parity measurements as native operations.
    The standard depolarizing noisy gate set (SD6) decomposes parity measurements into a reset, basis change rotations on the data qubits, two CNOTs, and a measurement.
    The operations can be pipelined such that rotating and interacting with the data qubits is the limiting factor.
    The superconducting inspired noisy gate set (SI1000) decomposes parity measurements into a reset, basis change rotations on the data and measurement qubits, two CZs, and a measurement.
    The reset and measurement are assumed to take much longer than the other operations, so the three edge measurement layers within a round are arranged to have their resets and measurements execute in parallel.
    }
    \label{fig:circuit_cycles}
\end{figure}

\begin{figure}[h]
    \centering
    \resizebox{0.8\linewidth}{!}{
        \includegraphics{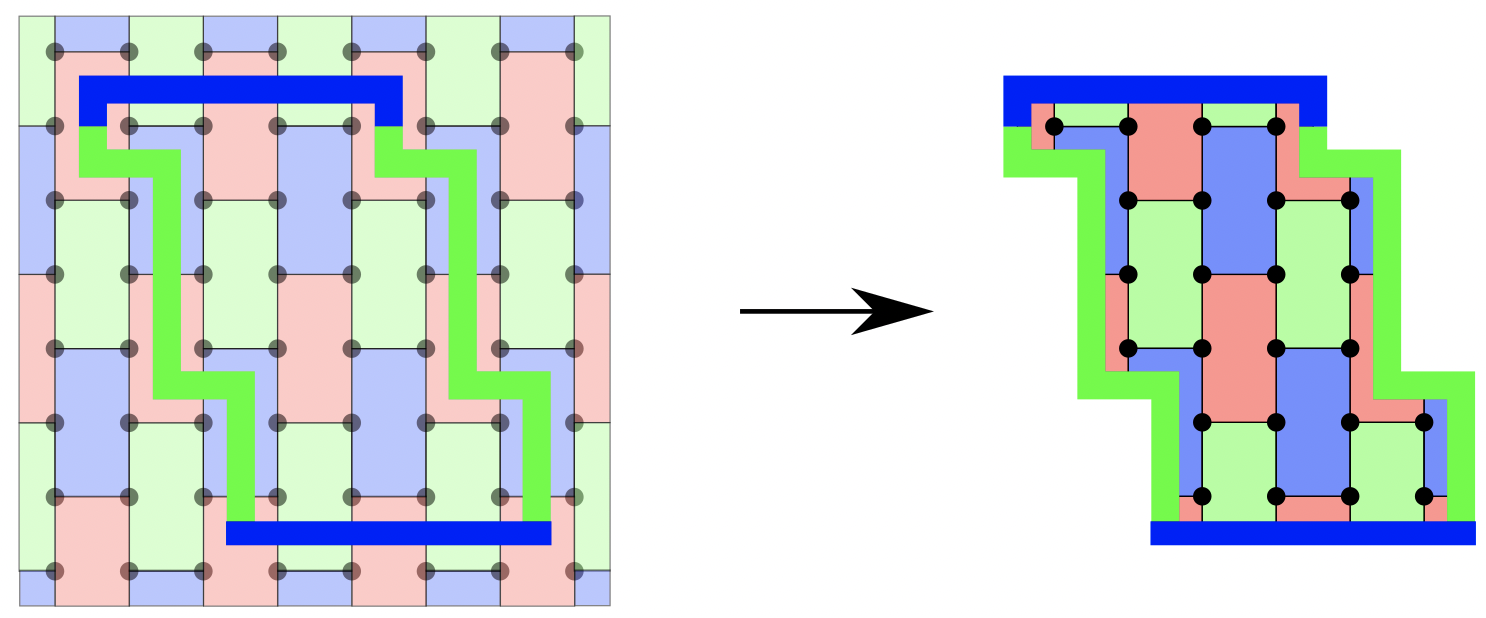}
    }
    \caption{
    A planar honeycomb code using a sheared patch shape.
    This was the initial patch shape we tried using, because shearing increases the minimum length of logical operators \cite{bombin2007optimal} in the periodic honeycomb code and because shearing makes the four boundaries symmetrical in the hexagonal lattice.
    But we found it has a suboptimal graphlike code distance due to the reduced diagonal distance between the two green boundaries.
    }
    \label{fig:sheared}
\end{figure}

\begin{figure}[h]
    \centering
    \resizebox{\linewidth}{!}{
        \includegraphics{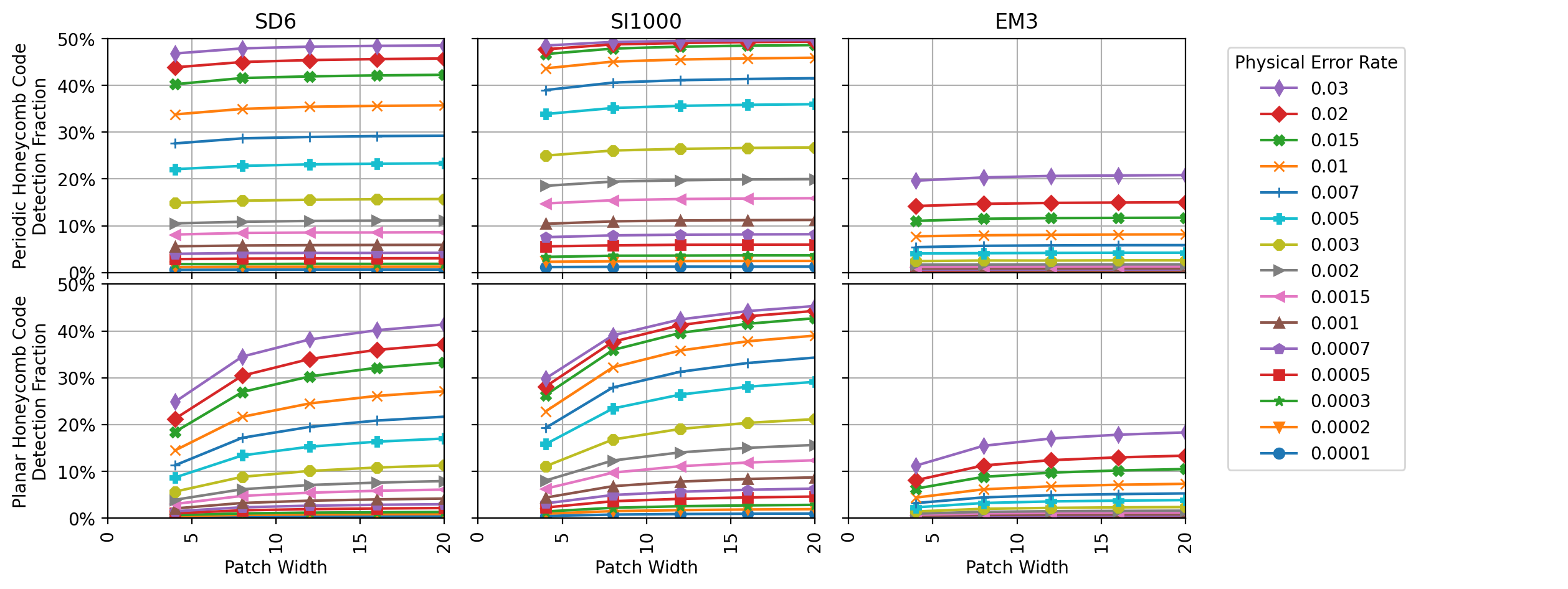}
    }
    \caption{
    Detection fractions of our previous implementation of the periodic honeycomb code and our new implementation of the planar honeycomb code.
    We expected the planar honeycomb code to have a higher overall detection fraction at large code distances, because of its alternating edge layer ordering, but the opposite appears to be the case.
    Raw grouped detection fraction data is in ancillary file ``detection\_fractions.csv".
    }
    \label{fig:detection_fractions}
\end{figure}

\begin{figure}[ht!]
    \centering
    \resizebox{\linewidth}{!}{
        \includegraphics{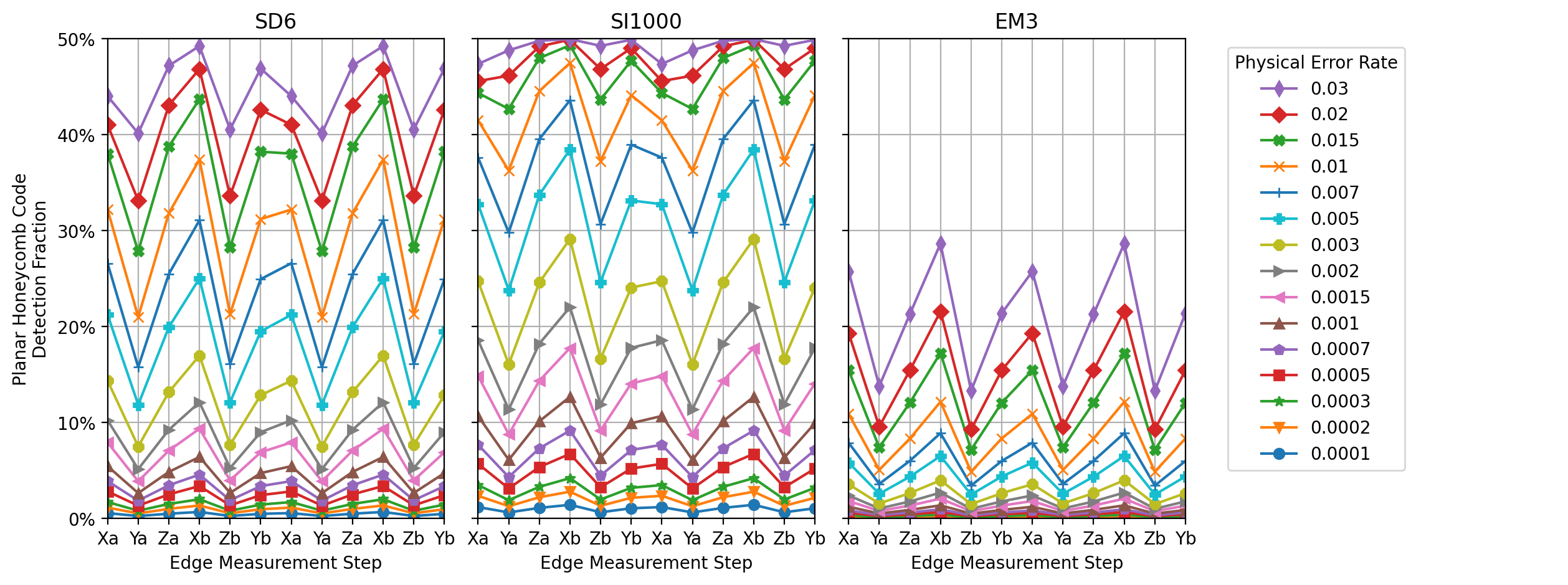}
    }
    \caption{
    Detection fractions grouped by edge measurement layer.
    The data is repeated (left to right) to help see the repeating cycle of the edge measurement layers.
    We expected these oscillations, because the alternating XYZ and XZY edge layer measurement orderings cause different parity checks to span different amounts of time and to aggregate different numbers of physical measurements.
    Raw grouped detection fraction data is in ancillary file ``detection\_fractions\_per\_layer.csv".
    }
    \label{fig:detection_fractions_over_time}
\end{figure}

\begin{figure}
    \centering
    \resizebox{\linewidth}{!}{
        \includegraphics{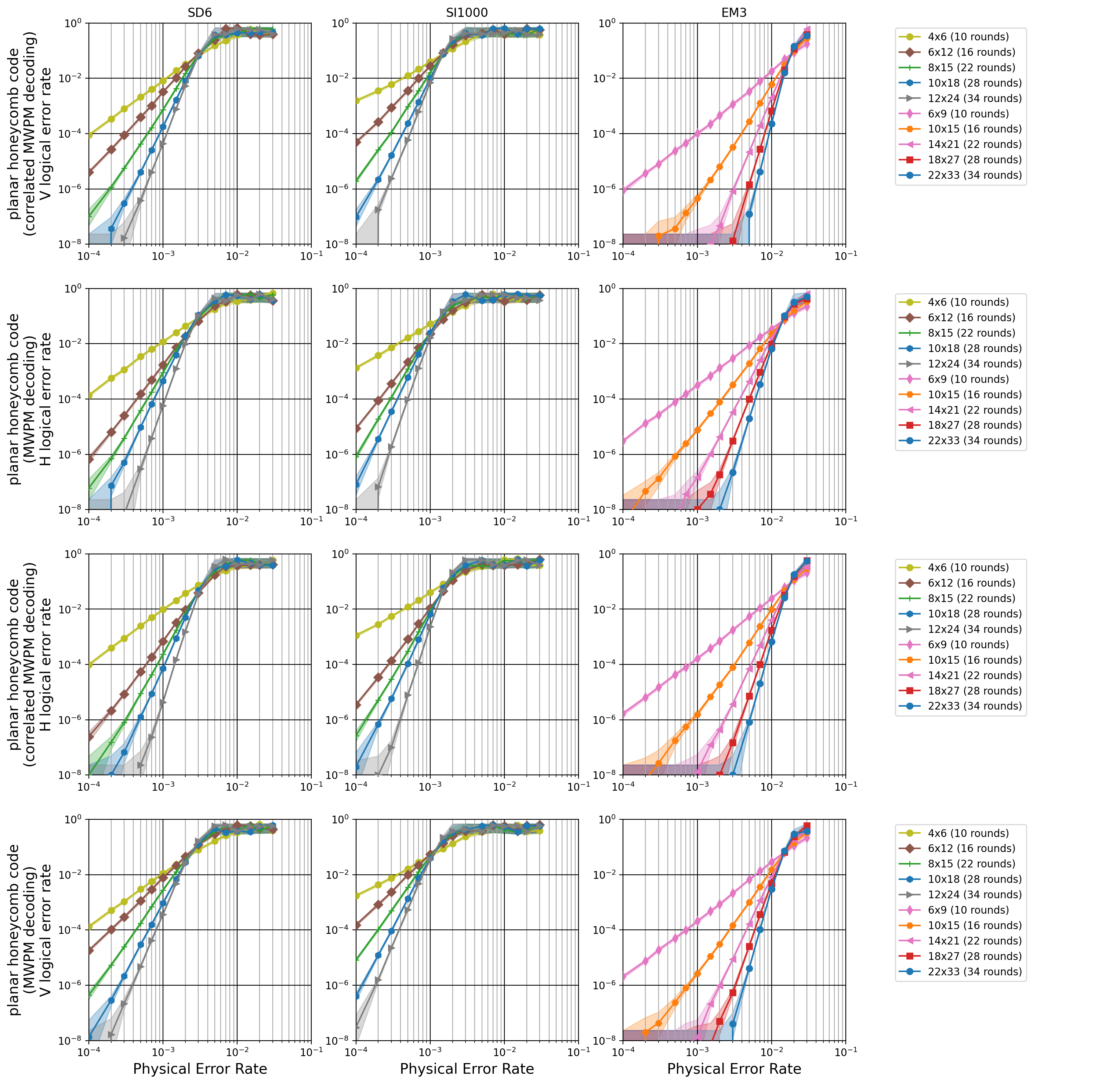}
    }
    \caption{
        Threshold plots for each decoder type, observable, and gate set.
        Raw logical error rate data is in ancillary file ``stats\_planar\_honeycomb.csv".
    }
    \label{fig:threshold_breakdown}
\end{figure}

\clearpage
\section{Hardware applicability of EM3 error models}
\label{app:hardware_em3}


Given the performance of the honeycomb code in the direct measurement model, the correspondence of the error model to hardware becomes an important question. Conveniently, the performance of the code in a direct measurement model is not especially sensitive to details of the error model, as we show here.

The EM3 error model in the main text was chosen for easy comparison to other literature, specifically \cite{chao2020optimization}. It doesn't correspond to the errors expected from a hardware implementation of two-qubit direct parity measurements. 

Here, we present two additional error models intended to mimic those provided for the standard circuit model. 
SDEM3 represents a simpler error model, with two-qubit depolarizing error applied after each two-qubit measurement gate, uncorrelated with a possible measurement flip.

SIEM3000 represents a more complex error model inspired by realizations of two-qubit parity measurements (MPP) in superconducting hardware. This error model applies two distinct error mechanisms with each MPP: first, a two-qubit error parallel to the measurement axis (e.g. a $ZZ$ error for MZZ), representing accidental measurement the individual qubit states rather than only their parity; and second, independent one-qubit errors perpendicular to the measurement (e.g. $X$ for MZZ) representing the independent likelihood of each qubit undergoing independent errors, such as $T_1$ decay. Note that this error model does not fully capture errors expected for a given hardware implementation of parity measurements and other more complex models might better reflect a spesific experimental realization. For example, one could imagine associating a $T_1$ decay error on one qubit during measurement with a bitflip on that qubit and an additional dephasing error on the other qubit. Rather than fully capturing possible hardware errors, these models are intended to serve as some evidence of the sensitivity of code performance in direct measurement models to the details of the error model used. \tbl{em3_mpp_noises} summarizes the re-definitions of MPP errors for these three models, with all other error rates being the same as for the EM3 model described in \tbl{models} and with individual errors given in \tbl{noise}. 

\begin{table}[h]
    \centering
    \resizebox{\linewidth}{!}{
    \begin{tabular}{|r|l|}
         \hline
         \textbf{Noise Model} & \textbf{MPP Error Implementation} \\
         \hline
         EM3 
         & Measure a Pauli product $PP$ on a pair of qubits and, with probability $p$, choose an error
         \\&uniformly from the set $\{I,X,Y,Z\}^{\otimes 2} \times \{\text{flip, no flip}\}$.
         \\&The flip error component causes the wrong result to be reported.
         \\&The Pauli error components are applied to the target qubits before the measurement.\\ 
         \hline
         SDEM3
         & Measure a Pauli product $PP$ on a pair of qubits,
         \\& and apply a two-qubit depolarizing channel with probability $p$: 
         \\& \begin{tabular}{l}
              Apply the trivial Pauli operator ($II$) with probability $1-p$
              \\  or any non-trivial Pauli operator with equal probability $p/15$\\
         \end{tabular}
         \\& and independently report the wrong measurement result with probability $p$\\
         \hline
         SIEM3000
         & Measure a Pauli product $PP$ on a pair of qubits,
         \\& and apply a two-qubit dephasing channel with probability $p$: 
         \\& \begin{tabular}{l}
              Apply the Pauli operator used for the measurement $PP$ with probability $p$
              \\ (e.g. for MZZ, apply ZZ with probability $p$)\\
         \end{tabular}
         \\& and, independently for each qubit, apply a bitflip error with probability $p$:
         \\& \begin{tabular}{l}
              Apply a Pauli operator $P' \neq P$ with probability $p$
              \\ (e.g. for MZZ, apply $X$ with probability $p$)\\
         \end{tabular}
         \\& and independently report the wrong measurement result with probability $p$\\ 
         \hline
         
    \end{tabular}
    }
    \caption{
        The MPP error mechanism for EM3, SDEM3 and SIEM3000.
        All other elements of the error model are identical to EM3 described in \tbl{models}, with elemental errors given in \tbl{noise}.
    }
    \label{tbl:em3_mpp_noises}
\end{table}

\begin{figure}
    \centering
    \resizebox{\linewidth}{!}{
        \includegraphics{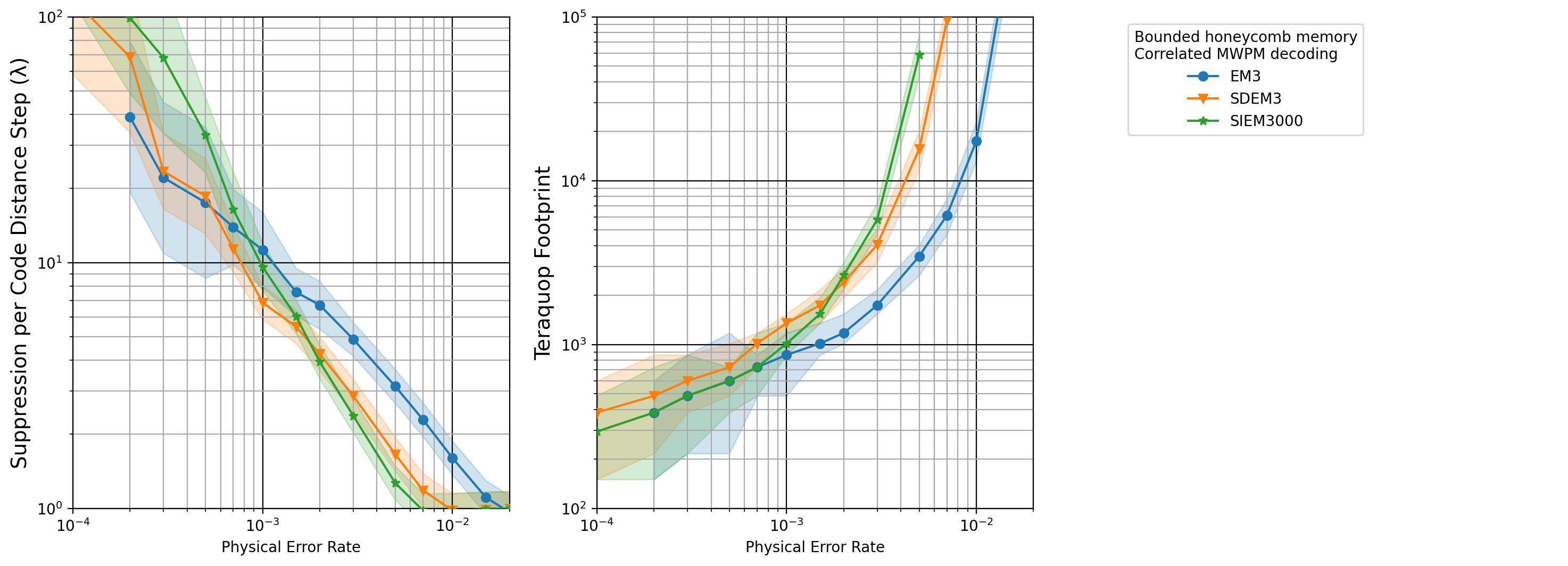}
    }
    \caption{
        Lambda factor and teraquop footprint plots for the three EM3-like error models described in \tbl{em3_mpp_noises}. EM3 is the error model used in the main text. SDEM3 is a more simplistic error model, without correlations between the 2-qubit depolarizing errors and the classical bit flip errors. SIEM3000 is a more complex model, representative of some errors expected in a hardware implementation. The latter two models have qualitatively similar performance, illustrating the some robustness to the precise details of the MPP error model.
    }
    \label{fig:em_comparison}
\end{figure}

\fig{em_comparison} illustrates the lambda factors and teraquop footprints for the three given error models. Specifically, the differences between the SDEM3 and SIEM3000 are of similar scale to the estimated uncertainty throughout the studied range. This demonstrated a relative insensitivity to the details of the error model.

All three error models coincide just above $\Lambda=10$, at an error parameter just below $p=10^{-3}$. Notably however, while SDEM3 and SIEM3000 have a similar thresholds below 1\%, the correlated EM3 model has noticeably higher threshold, and quite a distinct slope between threshold and the coincident point around $\Lambda=10$. This difference in scaling behaviour provides some reason to be cautious when quoting threshold alone as a summary of code performance under a given error model. Overall however, the qualitative agreement between these different error models for the MPP operation provides some evidence that the performance of the code is robust to changes in the error model, and is likely to remain similar for a true hardware error model for direct parity measurements. Details regarding the relative importance of different error mechanisms have been neglected here, and represent a good opportunity for future work, as does density matrix simulations of the errors associated with hardware parity measurements. 

\clearpage
\section{Example Planar Honeycomb Circuit}
\label{app:example_circuit}

The following is a Stim circuit file \cite{stimcircuitformat} describing a 4$\times$6 planar honeycomb memory experiment protecting the vertical observable for 100 rounds from ancillary file ``\path{example_planar_honeycomb_circuit.stim}".
Noise annotations have been removed so the circuit fits on the page.
In the EM3 error model the circuit would have a graphlike code distance of 2.

\begin{lstlisting}[style=stimcircuit]
QUBIT_COORDS(0, 0) 0
QUBIT_COORDS(0, 4) 1
QUBIT_COORDS(0, 5) 2
QUBIT_COORDS(1, 0) 3
QUBIT_COORDS(1, 1) 4
QUBIT_COORDS(1, 2) 5
QUBIT_COORDS(1, 3) 6
QUBIT_COORDS(1, 4) 7
QUBIT_COORDS(1, 5) 8
QUBIT_COORDS(2, 0) 9
QUBIT_COORDS(2, 1) 10
QUBIT_COORDS(2, 2) 11
QUBIT_COORDS(2, 3) 12
QUBIT_COORDS(2, 4) 13
QUBIT_COORDS(2, 5) 14
QUBIT_COORDS(3, 0) 15
QUBIT_COORDS(3, 1) 16
QUBIT_COORDS(3, 2) 17
QUBIT_COORDS(3, 3) 18
QUBIT_COORDS(3, 4) 19
QUBIT_COORDS(3, 5) 20
QUBIT_COORDS(4, 1) 21
QUBIT_COORDS(4, 2) 22
QUBIT_COORDS(4, 3) 23
# Transversal initialization.
R 0 1 2 3 4 5 6 7 8 9 10 11 12 13 14 15 16 17 18 19 20 21 22 23
TICK
H 0 1 2 3 4 5 6 7 8 9 10 11 12 13 14 15 16 17 18 19 20 21 22 23
TICK
# Round 1, layer Xa. Initial state had these edges as stabilizers.
MPP X1*X2 X4*X5 X8*X7 X11*X10 X13*X14 X16*X17 X20*X19 X22*X21 X0*X3 X6*X12 X9*X15 X18*X23
DETECTOR(0, 4.5, 0) rec[-12]
DETECTOR(1, 1.5, 0) rec[-11]
DETECTOR(1, 4.5, 0) rec[-10]
DETECTOR(2, 1.5, 0) rec[-9]
DETECTOR(2, 4.5, 0) rec[-8]
DETECTOR(3, 1.5, 0) rec[-7]
DETECTOR(3, 4.5, 0) rec[-6]
DETECTOR(4, 1.5, 0) rec[-5]
DETECTOR(0.5, 0, 0) rec[-4]
DETECTOR(1.5, 3, 0) rec[-3]
DETECTOR(2.5, 0, 0) rec[-2]
DETECTOR(3.5, 3, 0) rec[-1]
SHIFT_COORDS(0, 0, 1)
TICK
# Round 1, layer Ya. Get X*Y=Z faces for the first time.
MPP Y0 Y1 Y4*Y3 Y6*Y7 Y9*Y10 Y13*Y12 Y16*Y15 Y18*Y19 Y21 Y23 Y2 Y5 Y8*Y14 Y11*Y17 Y20 Y22
OBSERVABLE_INCLUDE(0) rec[-12] rec[-11]
SHIFT_COORDS(0, 0, 1)
TICK
# Round 1, layer Za. Get Y*Z=X faces. Initial state had them as stabilizers.
MPP Z0 Z2 Z3 Z6*Z5 Z8 Z9 Z11*Z12 Z14 Z15 Z18*Z17 Z20 Z22*Z23 Z1*Z7 Z4*Z10 Z13*Z19 Z16*Z21
OBSERVABLE_INCLUDE(0) rec[-11] rec[-10] rec[-9]
DETECTOR(2.5, 3, 0) rec[-27] rec[-25] rec[-19] rec[-10] rec[-7] rec[-2]
DETECTOR(1.5, 0, 0) rec[-30] rec[-28] rec[-14] rec[-11] rec[-3]
DETECTOR(1.5, 6, 0) rec[-20] rec[-12] rec[-9]
SHIFT_COORDS(0, 0, 1)
TICK
# Round 2, layer Xb. Get Z*X=Y faces for the first time.
MPP X1*X2 X4*X5 X8*X7 X11*X10 X13*X14 X16*X17 X20*X19 X22*X21 X0*X3 X6*X12 X9*X15 X18*X23
SHIFT_COORDS(0, 0, 1)
TICK
# Round 2, layer Zb. Get X*Z=Y faces again.
MPP Z0 Z2 Z3 Z6*Z5 Z8 Z9 Z11*Z12 Z14 Z15 Z18*Z17 Z20 Z22*Z23 Z1*Z7 Z4*Z10 Z13*Z19 Z16*Z21
OBSERVABLE_INCLUDE(0) rec[-11] rec[-10] rec[-9]
DETECTOR(1.5, 2, 0) rec[-41] rec[-38] rec[-31] rec[-13] rec[-10] rec[-3]
DETECTOR(3.5, 2, 0) rec[-35] rec[-33] rec[-29] rec[-7] rec[-5] rec[-1]
DETECTOR(0.5, 5, 0) rec[-43] rec[-40] rec[-32] rec[-15] rec[-12] rec[-4]
DETECTOR(2.5, 5, 0) rec[-37] rec[-34] rec[-30] rec[-9] rec[-6] rec[-2]
DETECTOR(2.5, -1, 0) rec[-39] rec[-36] rec[-11] rec[-8]
DETECTOR(0.5, -1, 0) rec[-44] rec[-42] rec[-16] rec[-14]
SHIFT_COORDS(0, 0, 1)
TICK
# Round 2, layer Yb. Get Z*Y=X faces again.
MPP Y0 Y1 Y4*Y3 Y6*Y7 Y9*Y10 Y13*Y12 Y16*Y15 Y18*Y19 Y21 Y23 Y2 Y5 Y8*Y14 Y11*Y17 Y20 Y22
OBSERVABLE_INCLUDE(0) rec[-12] rec[-11]
DETECTOR(2.5, 3, 0) rec[-71] rec[-69] rec[-63] rec[-54] rec[-51] rec[-46] rec[-26] rec[-23] rec[-18] rec[-11] rec[-9] rec[-3]
DETECTOR(4.5, 3, 0) rec[-67] rec[-61] rec[-49] rec[-21] rec[-7] rec[-1]
DETECTOR(0.5, 3, 0) rec[-75] rec[-73] rec[-65] rec[-57] rec[-48] rec[-29] rec[-20] rec[-15] rec[-13] rec[-5]
SHIFT_COORDS(0, 0, 1)
TICK

# Stable state reached. Can now consistently compare to stabilizers from previous rounds.
REPEAT 48 {
    # Layer Xa.
    MPP X1*X2 X4*X5 X8*X7 X11*X10 X13*X14 X16*X17 X20*X19 X22*X21 X0*X3 X6*X12 X9*X15 X18*X23
    DETECTOR(1.5, 4, 0) rec[-98] rec[-96] rec[-91] rec[-85] rec[-83] rec[-76] rec[-25] rec[-23] rec[-16] rec[-10] rec[-8] rec[-3]
    DETECTOR(2.5, 1, 0) rec[-97] rec[-95] rec[-90] rec[-84] rec[-82] rec[-75] rec[-24] rec[-22] rec[-15] rec[-9] rec[-7] rec[-2]
    SHIFT_COORDS(0, 0, 1)
    TICK
    # Layer Ya.
    MPP Y0 Y1 Y4*Y3 Y6*Y7 Y9*Y10 Y13*Y12 Y16*Y15 Y18*Y19 Y21 Y23 Y2 Y5 Y8*Y14 Y11*Y17 Y20 Y22
    OBSERVABLE_INCLUDE(0) rec[-12] rec[-11]
    DETECTOR(1.5, 4, 0) rec[-41] rec[-39] rec[-32] rec[-13] rec[-11] rec[-4]
    DETECTOR(2.5, 1, 0) rec[-40] rec[-38] rec[-31] rec[-12] rec[-10] rec[-3]
    DETECTOR(4.5, 1, 0) rec[-36] rec[-29] rec[-8] rec[-1]
    DETECTOR(-0.5, 4, 0) rec[-43] rec[-34] rec[-15] rec[-6]
    DETECTOR(0.5, 1, 0) rec[-44] rec[-42] rec[-33] rec[-16] rec[-14] rec[-5]
    DETECTOR(3.5, 4, 0) rec[-37] rec[-35] rec[-30] rec[-9] rec[-7] rec[-2]
    SHIFT_COORDS(0, 0, 1)
    TICK
    # Layer Za.
    MPP Z0 Z2 Z3 Z6*Z5 Z8 Z9 Z11*Z12 Z14 Z15 Z18*Z17 Z20 Z22*Z23 Z1*Z7 Z4*Z10 Z13*Z19 Z16*Z21
    OBSERVABLE_INCLUDE(0) rec[-11] rec[-10] rec[-9]
    DETECTOR(2.5, 3, 0) rec[-70] rec[-67] rec[-62] rec[-55] rec[-53] rec[-47] rec[-27] rec[-25] rec[-19] rec[-10] rec[-7] rec[-2]
    DETECTOR(1.5, 0, 0) rec[-74] rec[-71] rec[-63] rec[-58] rec[-56] rec[-30] rec[-28] rec[-14] rec[-11] rec[-3]
    DETECTOR(1.5, 6, 0) rec[-72] rec[-69] rec[-48] rec[-20] rec[-12] rec[-9]
    SHIFT_COORDS(0, 0, 1)
    TICK
    # Layer Xb.
    MPP X1*X2 X4*X5 X8*X7 X11*X10 X13*X14 X16*X17 X20*X19 X22*X21 X0*X3 X6*X12 X9*X15 X18*X23
    DETECTOR(1.5, 2, 0) rec[-99] rec[-97] rec[-91] rec[-85] rec[-82] rec[-75] rec[-25] rec[-22] rec[-15] rec[-11] rec[-9] rec[-3]
    DETECTOR(3.5, 2, 0) rec[-95] rec[-93] rec[-89] rec[-79] rec[-77] rec[-73] rec[-19] rec[-17] rec[-13] rec[-7] rec[-5] rec[-1]
    SHIFT_COORDS(0, 0, 1)
    TICK
    # Layer Zb.
    MPP Z0 Z2 Z3 Z6*Z5 Z8 Z9 Z11*Z12 Z14 Z15 Z18*Z17 Z20 Z22*Z23 Z1*Z7 Z4*Z10 Z13*Z19 Z16*Z21
    OBSERVABLE_INCLUDE(0) rec[-11] rec[-10] rec[-9]
    DETECTOR(1.5, 2, 0) rec[-41] rec[-38] rec[-31] rec[-13] rec[-10] rec[-3]
    DETECTOR(3.5, 2, 0) rec[-35] rec[-33] rec[-29] rec[-7] rec[-5] rec[-1]
    DETECTOR(0.5, 5, 0) rec[-43] rec[-40] rec[-32] rec[-15] rec[-12] rec[-4]
    DETECTOR(2.5, 5, 0) rec[-37] rec[-34] rec[-30] rec[-9] rec[-6] rec[-2]
    DETECTOR(2.5, -1, 0) rec[-39] rec[-36] rec[-11] rec[-8]
    DETECTOR(0.5, -1, 0) rec[-44] rec[-42] rec[-16] rec[-14]
    SHIFT_COORDS(0, 0, 1)
    TICK
    # Layer Yb.
    MPP Y0 Y1 Y4*Y3 Y6*Y7 Y9*Y10 Y13*Y12 Y16*Y15 Y18*Y19 Y21 Y23 Y2 Y5 Y8*Y14 Y11*Y17 Y20 Y22
    OBSERVABLE_INCLUDE(0) rec[-12] rec[-11]
    DETECTOR(2.5, 3, 0) rec[-71] rec[-69] rec[-63] rec[-54] rec[-51] rec[-46] rec[-26] rec[-23] rec[-18] rec[-11] rec[-9] rec[-3]
    DETECTOR(4.5, 3, 0) rec[-67] rec[-61] rec[-49] rec[-21] rec[-7] rec[-1]
    DETECTOR(0.5, 3, 0) rec[-75] rec[-73] rec[-65] rec[-57] rec[-48] rec[-29] rec[-20] rec[-15] rec[-13] rec[-5]
    SHIFT_COORDS(0, 0, 1)
    TICK
}

# Transversal measurement.
MPP X0 X1 X2 X3 X4 X5 X6 X7 X8 X9 X10 X11 X12 X13 X14 X15 X16 X17 X18 X19 X20 X21 X22 X23
# Got X faces again.
DETECTOR(2.5, 3, 1) rec[-50] rec[-47] rec[-42] rec[-35] rec[-33] rec[-27] rec[-13] rec[-12] rec[-11] rec[-7] rec[-6] rec[-5]
DETECTOR(1.5, 0, 1) rec[-54] rec[-51] rec[-43] rec[-38] rec[-36] rec[-21] rec[-20] rec[-15] rec[-14]
DETECTOR(1.5, 6, 1) rec[-52] rec[-49] rec[-28] rec[-16] rec[-10]
# Got Y*X = Z faces again.
DETECTOR(1.5, 4, 0) rec[-110] rec[-108] rec[-103] rec[-97] rec[-95] rec[-88] rec[-37] rec[-35] rec[-28] rec[-18] rec[-17] rec[-16] rec[-12] rec[-11] rec[-10]
DETECTOR(2.5, 1, 0) rec[-109] rec[-107] rec[-102] rec[-96] rec[-94] rec[-87] rec[-36] rec[-34] rec[-27] rec[-15] rec[-14] rec[-13] rec[-9] rec[-8] rec[-7]
OBSERVABLE_INCLUDE(0) rec[-15] rec[-14] rec[-12] rec[-11]
\end{lstlisting}

\end{document}